\documentclass[%
 reprint,
nofootinbib,
 amsmath,amssymb,
 aps,
 onecolumn,
]{revtex4-2}

\usepackage{graphicx}% Include figure files
\usepackage{dcolumn}% Align table columns on decimal point
\usepackage{bm}% bold math
\usepackage{color}

\usepackage[backgroundcolor=gray!30,linecolor=black]{todonotes}
\usepackage{amsfonts, amssymb, amsmath, amsthm, mathrsfs, bbm, cjhebrew, gensymb, textcomp, mathtools, dsfont, calligra, physics}

\usepackage{caption}
\usepackage{subcaption}

\newtheorem{remark}{Remark}

\newtheorem{Thm}{Theorem}[section]

\newtheorem{Rmk}[Thm]{Remark}

\newtheorem*{Thm*}{Theorem}

\usepackage[normalem]{ulem} % For \sout (strike out)
\usepackage{xcolor}

\usepackage{placeins} % Used only for \FloatBarrier

\begin{document}

\preprint{APS/123-QED}

\title{Identifying the body force from partial observations of a 2D incompressible velocity field}
%\\Relax, then punch:\\Recovering the forcing function from partial observations of a given velocity field}% Force line breaks with \\

\author{Aseel Farhat}
\email{afarhat@fsu.edu}
\affiliation{Florida State University\\
Department of Mathematics}
\author{Adam Larios}%
 \email{alarios@unl.edu}
\affiliation{University of Nebraska\\
Department of Mathematics
}%


\author{Vincent R. Martinez}
 \email{vrmartinez@hunter.cuny.edu}
\affiliation{Hunter College, City University of New York\\
Department of Mathematics \& Statistics\\
Graduate Center, City University of New York\\
Department of Mathematics
}%

%\author{Benjamin Pachev}
%\email{benjaminpachev@utexas.edu}
%\affiliation{University of Texas at Austin\\
%Oden Institute for Computational Science and Engineering}%

\author{Jared P. Whitehead}
\email{whitehead@mathematics.byu.edu}
\affiliation{Brigham Young University\\
Department of Mathematics
}%



\begin{abstract}
Using limited observations of the velocity field of the two-dimensional  Navier-Stokes equations, we successfully reconstruct the steady body force that drives the flow.  The number of observed data points is less than 10\% of the number of modes that describes the full flow field, indicating that the method introduced here is capable of identifying complicated forcing mechanisms from a relatively small collection of observations.  In addition to demonstrating the efficacy of this method on turbulent flow data generated by simulations of the two-dimensional Navier-Stokes equations, we also rigorously justify convergence of the derived algorithm.  Beyond the practical applicability of such an algorithm, the reliance of this method on the dynamical evolution of the system  yields physical insight into the turbulent cascade.
%\begin{description}
%item[Structure]
%\end{description}
%{\color{red}\bf V: Color for Vincent's edits}
%\adam{A: Color for Adam's edits}
%\adam{A: Note: We somehow cite the present work, ``Conjuring the force...''.  It appears in the bibliography, but I can't seem to find where we reference it.}
\end{abstract}

\maketitle

%\tableofcontents
\noindent
\section{Introduction}
\noindent
In an ancient dusty magic school, where wizards cast spells by moving their hands in complex arcane patterns, a young student, Henri Pother, 
% Note: ``Henri'' is for Henri Poincare due to the tie-in with dynamical systems.  ``Pother'' means a commotion, or a choking cloud of smoke, dust, etc.  Of course, this is a pretty explicit stand-in for Harry Potter.
hides behind a curtain watching a master wizard perform his most secret spell.  Unfortunately for Henri, he can only see through a few small holes in the curtain.  To make matters worse, the wizard is invisible!  All Henri can see is the motion of the dust in the air around the wizard's invisible hands.  \textit{Can Henri reconstruct the movement of the wizard's hands from only limited observations of the air currents?}

The above make-believe scenario is an analogue for the real-world setting of weather modeling: We have equations that model weather, but we do not know how these equations are forced (we don't know the movement of the wizard's hands). On the other hand, we observe parts of the evolution on large-scales (the motion of the dust particles from which one can systematically construct an approximation of the surrounding air current velocity, for instance through Particle Image Velocimetry \cite{WillertGharib1991}). The real question that we want to answer is then: \textit{Can we systematically reconstruct the forcing on the system from only sparse observations of the velocity?}  In the present paper, we show that the answer is ``yes,'' at least in the context of the 2D incompressible Navier-Stokes equations with periodic boundary conditions.\footnote{We consider periodic boundary conditions here as a proof-of-concept, but there seems to be no fundamental computational difficulty in adapting this algorithm to a wide variety of practical settings, including the 3D case with physical boundary conditions.  This slightly more nuanced and computationally intensive setting will be investigated in a future work.} This is accomplished through an algorithm that enforces exponential decay of state errors on observational scales.  We mathematically prove that that this algorithm reconstructs non-potential driving forces that are external to the system, that is, state-independent. We probe the practical efficacy of this algorithm in a scenario of turbulent flow. The reference flow field is generated by a highly resolved numerical solution of the 2D Navier-Stokes equation with a randomized low-mode force to which the algorithm is agnostic. When state observations are made across all directly-forced scales, we observe both model and state error to achieve machine-precision convergence to the true values exponentially fast in time.

Adequate control of fluid flow is a difficult problem that has been extensively studied (see, e.g., \cite{gunzburger2002perspectives,gunzburger2012flow}) particularly in light of modern advances in computing. Motivation for such studies is found in a number of disciplines in the physical and engineering sciences,  where the ability to control either classical Newtonian fluids or complex non-Newtonian fluids is of significant interest.
In \cite{Azouani_Olson_Titi_2014} a feedback control mechanism (motivated byt the original work of \cite{hoke1976initialization}) was exploited to smoothly merge incoming partial observations with a dynamic model given by a system of partial differential equations that is simultaneously evolved forward in time, allowing for a controlled estimate of the true full state of the unobserved system.  This is often referred to as the Azouani-Olson-Titi (AOT) algorithm or Continuous Data Assimilation (CDA).  The AOT approach has by now been studied in many fluid systems, yielding both numerical evidence of its efficacy and rigorous analysis to justify the convergence of the generated approximating state to its true value.  The current investigation exploits this feedback control approach to produce a rigorously justified approach to the determination of a steady, but apriori unknown, external driving force from partial observations of the flow field in two-dimensional turbulence.

A common assumption found in many of the works that incorporate feedback control in fluid systems is the existence and complete knowledge of a perfect model for the dynamical evolution of the fluid itself, that is, the underlying physical model is known exactly beforehand.  Notable exceptions to this framework are the studies performed in \cite{FarhatGlattHoltzMartinezMcQuarrieWhitehead2020} and \cite{Carlson_Hudson_Larios_2020}, where the true dynamic model is known up to a set of unidentified scalar parameters.  In \cite{FarhatGlattHoltzMartinezMcQuarrieWhitehead2020}, the canonical Rayleigh-B\'enard convection setting is investigated where the exact value of the Prandtl number (the nondimensional quantity denoting the ratio of kinematic viscosity to thermal diffusivity) is unknown.  Observations extracted from an identical setup, but with a different value of the Prandtl number are then rigorously shown to drive the simulated solution toward the true state up to a constant error that is Prandtl number-dependent. This analysis is then supplemented with a battery of numerical simulations that verify the mathematical conclusions and probe the efficacy of the method by exploring situations outside of the regimes where convergence can mathematically be guaranteed. In \cite{Carlson_Hudson_Larios_2020}, a similar study is performed on the periodic 2D Navier-Stokes equations with an unknown viscosity, but a further advancement is introduced which exploits the numerical observation that the state error relaxes to be relatively constant.  This remaining state error is then used to propose an approximate value to the true viscosity; this procedure can then be iterated to produce a sequence of approximating values that reliably converge to the true viscosity value up to numerical precision. In a series of recent papers \cite{Carlson_Hudson_Larios_Martinez_Ng_Whitehead_2021, martinez2022convergence, martinez2022force}, the underlying mechanisms leading to the observed convergence were identified in a mathematically salient way to supply an analytical proof of this convergence. The current investigation takes this a step further, providing both rigorous justification and numerical evidence for an algorithm capable of recovering the entire inhomogeneous term in a fluid system, which in contrast to inferring a single parameter such as viscosity, entails the inference of a potentially large number of parameters at once.

A variation to the approach for parameter recovery that was originally introduced in \cite{Carlson_Hudson_Larios_2020} was proposed in \cite{pachev2022concurrent}. There, a more principled perspective based on enforcing a form of null-controllability was taken to develop a continuously-updating parameter algorithm in contrast to the sequentially-updating algorithm in \cite{Carlson_Hudson_Larios_2020}. This variation was numerically studied in the context of the one-dimensional Kuramoto-Sivashinsky equation and was seen to reliably infer multiple unknown parameters appearing in the system in a concurrent fashion.  The algorithm of interest in the present study is primarily based on the one introduced in \cite{pachev2022concurrent}. In contrast to \cite{pachev2022concurrent}, however, the current article, firstly, negotiates a situation that possesses richer dynamical behavior in two-dimensional turbulent flow, in addition to one that contains significantly more parameters to infer in having to determine all coefficients of the forcing function, the number of which may potentially be very large, and secondly, supplies the first rigorous proof of convergence of a parameter recovery algorithm that is based on the null-controllability perspective from \cite{pachev2022concurrent}. 

At this point, we remark on three particular recent works that directly relate to the task of reconstructing driving forces in a system, namely \cite{FarhatJollyTiti2015, AltafTitiGebraelKnioZhaoMcCabeHoteit2017, AgasthyaDiLeoniBiferale2022} and \cite{martinez2022force}. In the first set of references, the problem of inferring temperature from velocity measurements \cite{FarhatJollyTiti2015, AltafTitiGebraelKnioZhaoMcCabeHoteit2017} or velocity from temperature measurements \cite{AltafTitiGebraelKnioZhaoMcCabeHoteit2017, AgasthyaDiLeoniBiferale2022} in the context of Rayleigh-B\'enard convection (RBC) is studied. In \cite{FarhatJollyTiti2015}, a downscaling algorithm for recovering temperature from velocity measurements in the planar setting (2D) is analytically proven to converge \cite{FarhatJollyTiti2015} and its efficacy was studied in several numerical tests in \cite{AltafTitiGebraelKnioZhaoMcCabeHoteit2017}. In \cite{AltafTitiGebraelKnioZhaoMcCabeHoteit2017, AgasthyaDiLeoniBiferale2022} the more difficult case of inferring velocity from temperature measurements was also studied numerically with mixed results; the former carried out numerical experiments in the 2D setting and there identified a possible mechanism for generating asynchrony between the approximating velocity field and the reference velocity field through a Kelvin-Helmholtz instabilty, while the latter carried out their experiments in the 3D setting and identified more optimistic scenarios. In contrast to the present article, recovery of the temperature in RBC may be viewed as a form of forcing recovery. However, one notable and very important difference between \cite{AgasthyaDiLeoniBiferale2022} and the present article is that in the setting of RBC, the driving force (temperature) present in the velocity field is linearly coupled to the velocity, which the analysis in \cite{FarhatJollyTiti2015} relies on in a crucial way. In the situation studied here, the driving force is completely external to the system and no relation whatsoever between the state and the driving force is posited apriori. Indeed, the success of the algorithm studied here is independent of the existence of any relation between the state and forcing.

In \cite{martinez2022force}, the convergence properties are studied of an algorithm for inferring $\mathbf{f}$ in \eqref{eq:NS_true} that is similar to the one treated in the current article. However, in \cite{martinez2022force}, the state variable is reconstructed by directly using the observations as the low-mode approximation and subsequently generating the unobserved high-modes with an incorrect forcing via a nudging-based method. A correct forcing is then iteratively generated by exploiting the equation itself to systematically filter errors. In contrast, the algorithm studied in the current article generates an approximation to the state variable by directly enforcing exponential convergence of the \textit{state error} at the observational scale. As mentioned above, this idea was initially introduced in \cite{pachev2022concurrent} in the context of the one-dimensional Kuramoto-Sivashinsky equation. From this point of view, we are able to supply a unifying perspective between the two algorithms obtained by different derivations, that distinguishes each algorithm within a family of variations of one common approach. We refer the reader to Remark \ref{rmk:algorithms} for further details.  Another notable remark is that the work of \cite{martinez2022force} only provided theoretical justification of the algorithm studied therein and did not provide numerical confirmation of its effectiveness. In this article, we carry out both theoretical and numerical studies of the proposed algorithm. In addition to identifying a theoretical framework in which convergence of the algorithm can be rigorously guaranteed, we moreover robustly observe convergence of the proposed algorithm up to the level of machine precision. It is noteworthy that we were unable to produce comparable numerical results for the algorithm proposed in \cite{martinez2022force}, perhaps suggesting that the approach developed here may possess fundamental practical advantages over the one developed in \cite{martinez2022force}. Overall, the current article complements the works referenced above firstly, by proposing a new algorithm that continuously processes observations for reconstructing the force, secondly, by rigorously demonstrating its convergence via mathematical proof, and thirdly, providing substantive numerical evidence of its efficacy and robustness.

Although we focus on the idealized test-case of 2D isotropic turbulence, we note that the methods presented here can be readily adapted to other dynamical systems, such as the Magnetohydrodynamic (MHD) equations, the Boussinesq equations of ocean flow, the setting of Rayleigh-Benard convection, geophysical flows such as the surface quasi-geostrophic (SQG) equations or the primitive equations of the ocean, pattern formation equations such as the Cahn-Hilliard or Allen-Cahn equations, equations of flame fronts and crystal growth, such as the Kuramoto-Sivashisnky equation, and many others.  In addition, it is straight-forward (at least, at the algorithmic level) to adapt the methods presented here to the setting of physical boundaries.  The 3D case could also be considered, although with the usual caveats regarding the analytical difficulties of the 3D Navier-Stokes equations (see, e.g., \cite{Biswas_Price_2020_AOT3D} for some results on the AOT algorithm in 3D).  

Lastly, we remark that the approaches described above for discovery of model parameters while simultaneously reconstructing the true state of the system is readily comparable to other data-driven model recovery techniques such as SINDy (see \cite{brunton2016discovering}, for example, or \cite{djeumou2022learning,mojgani2022discovery} and the references therein).  One particular benefit of the feedback control mechanism exploited here for the purpose of recovering the unknown forcing function in 2D turbulence, is that the parameter estimation is performed \textit{on-the-fly} in the fashion of continuous data assimilation and without any need for post-processing. Indeed, the data is inserted dynamically into the model in real-time, which is, in turn, incrementally updated to obtain the correct parameters that characterize the external forcing.

The rest of this article is organized as follows: Section \ref{sec:pre} will establish notation and the necessary preliminary results required by the analysis that follows.  Section \ref{sec:pde} provides a formal derivation of the forcing update algorithm.  Section \ref{sec:analysis} provides the rigorous analysis that justifies the algorithm in 2D.  Section \ref{sec:simulations} describes the numerical simulations that were performed that demonstrate how well the algorithm performs in 2D, and Section \ref{sec:conclusions} concludes with a takeaway message and potential for future work.  A reader that is less interested in the technical details of the mathematical analysis may be well-served to focus on Sections \ref{sec:pde}, \ref{sec:simulations} and \ref{sec:conclusions}.

%%%%%%%%%%%%%%%%%%%%%%%%%%%%%%%%%%%%%%%%%%%%%%%%
\section{The equations of motion and their mathematical setting} \label{sec:pre}
%%%%%%%%%%%%%%%%%%%%%%%%%%%%%%%%%%%%%%%%%%%%%%%%
In this section, we provide a short description of the mathematical model of interest and the functional setting in which we carry out our convergence analysis.  The development of the underlying algorithm in Section \ref{sec:pde} can be followed without some of the definitions provided here, but these definitions and preliminary results are necessary for the rigorous analysis performed in Section \ref{sec:analysis} that completely justifies the algorithm.

We consider an incompressible fluid in a periodic domain, $\Omega=[0,L]^d$, of length $L>0$ (all of the analysis and simulations presented below assume that $d=2$, but the heuristic derivation of the algorithm is independent of the dimension $d$), subject to a time-independent external body force $\mathbf{f}$, which is mean-free over $\Omega$. The evolution of the velocity field is governed by the Navier-Stokes equations given by
\begin{equation}\label{eq:NS_true}
    \partial_t\mathbf{u} + \mathbf{u}\cdot \nabla \mathbf{u} =-\nabla p+ \nu \nabla^2 \mathbf{u} + \mathbf{f},\quad \nabla\cdot\mathbf{u} = 0,
\end{equation}
where fluid density has been normalized to unity, the kinematic viscosity, $\nu$, is known, and $\mathbf{u}, p$ are assumed mean-free over $\Omega$. We assume that the body force is unknown and that the velocity field is partially observed. The main objective is to reconstruct the non-conservative component of $\mathbf{f}$ at \textit{observational scales} by leveraging the model jointly with the observations. For convenience, let us therefore assume that $\nabla\cdotp\mathbf{f}=0$.

An important non-dimensional quantity that serves as a proxy for the Reynolds number when the Navier-Stokes equations is externally forced is the Grashof number, which we will denote by $G$. In two-dimensions, $G$ is defined by
\begin{align}\label{Grashof}
G = \frac{\norm{\bf f}_{2}}{(\kappa_0\nu)^2},\quad \text{where}\  \kappa_0:=\frac{2\pi}{L},
\end{align}
where $\|\hspace{1pt}\cdotp\|_2$ is the $L^2$ norm over $\Omega$.  Note that $\kappa_0$ is the smallest eigenvalue of the negative Laplacian $-\Delta$, while in the context of three-dimensional turbulence, the Grashof number is known to be related to the Reynolds number, $\text{Re}$, of the flow as $G\sim\text{Re}^2$ (see  \cite{DascaliucFoiasJolly2009}).
For our purposes, we will assume that the body force is twice weakly differentiable over $\Omega$, all of whose weak partial derivatives are square-integrable over $\Omega$. It is known that the corresponding initial value problem for the 2D NSE \eqref{eq:NS_true} is well-posed in this setting and, moreover, the dynamics for the system possesses a finite-dimensional global attractor when ${\bf f}$ is time-independent, see e.g., \cite{ConstantinFoias88, Robinson2003, Temam1997}; additional properties of the solution which are relevant to the analysis performed later in the article is provided in Section \ref{sec:appendix}.

%%%%%%%%%%%%%%%%%%%%%%%%%%%%%%%%%%%%%%%%%%%%%%%%%
\section{Derivation of the algorithm and Heuristics for convergence}\label{sec:pde}
%%%%%%%%%%%%%%%%%%%%%%%%%%%%%%%%%%%%%%%%%%%%%%%%
The method for parameter recovery studied in this paper makes crucial use of a state-recovery algorithm originally introduced by Azouani, Olson, and Titi (AOT) \cite{Azouani_Olson_Titi_2014} in the context of continuous data assimilation for the 2D NSE equations; the system associated to this method is introduced below in \eqref{eq:NS_nudge}. In the seminal work of AOT, a {\it feedback control paradigm} is introduced for obtaining an approximation of the state using a \textit{sufficiently large, but finite number of observations} of the true solution, collected continuously in time.  This approximation asymptotically converges to the true solution corresponding to the observed data at an exponential rate. This particular algorithm differed from previously studied algorithms through the manner in which observations were assimilated; in \cite{Azouani_Olson_Titi_2014} observations were inserted into the dynamical model as an external forcing term that enforced relaxation to the true state of the system, whereas previous algorithms such as that studied in \cite{OlsonTiti03} had inserted observations into the model by directly replacing dynamical terms. The use of such schemes for numerical weather prediction has been and continues to be extensively studied (see \cite{AbarbanelShirmanBreenKadakiaReyArmstrongMargoliash2017, Baek2007, SzendroRodriguezLopez2009, YangBakerLiCordesHuffNagpalOkerekeVillafaneKalnayDuane2006} for instance).  
Since its introduction to the mathematical fluid dynamics community, this algorithm has been a topic of intense activity (see \cite{AlbanezBenvenutti18,
AlbanezLopesTiti2016,
AltafTitiGebraelKnioZhaoMcCabeHoteit2017,
BessaihOlsonTiti2015,
BiswasFoiasMondainiTiti2019,
Biswas_Hudson_Larios_Pei_2017,
BiswasBrownMartinez2022,
BiswasMartinez2017,
BlocherMartinezOlson2018,
BlomkerLawStuartZygalakis2013,
BrettLamLawMcCormickScottStuart2013,
Carlson_Hudson_Larios_2020,
Carlson_Hudson_Larios_Martinez_Ng_Whitehead_2021,
Carlson_Larios_Titi_2023_nlDA,
Carlson_VanRoekel_Petersen_Godinez_Larios_2021,
KnioDesamsettiDasariLangodanHoteitTiti2019,
FarhatJohnstonJollyTiti2018,
FarhatJollyTiti2015,
FarhatLunasinTiti17,
FarhatLunasinTiti2016a,
FarhatLunasinTiti2016b,
FarhatLunasinTiti2016c,
FarhatLunasinTiti2017,
FoiasMondainiTiti2016,
Franz_Larios_Victor_2022,
Gardner_Larios_Rebholz_Vargun_Zerfas_2020_VVDA,
GeshoOlsonTiti2016,
IbdahMondainiTiti2020,
JollyMartinezOlsonTiti2019,
JollyMartinezTiti2017,
jolly2021data,
JollySadigovTiti2015,
JollySadigovTiti2017,
Larios_Pei_2017_KSE_DA_NL,
Larios_Pei_2018_NSV_DA,
Larios_Pei_Victor_2023_second_best,
Larios_Petersen_Victor_2023,
Larios_Rebholz_Zerfas_2018,
Larios_Victor_2019,
Larios_Victor_2023_fracAOT,
LeoniMazzinoBiferale2018,
LeoniMazzinoBiferale2020,
MarkowichTitiTrabelsi2016,
MondainiTiti18}), and can be considered a nonlinear complement to other data assimilation methods such as variations on the Kalman filter (see e.g., \cite{tandeo2020review}), variational methods, particle filtering, and several other related methods (see \cite{YangCorazzaCarrassiKalnayMiyoshi2009,Poterjoy2016,CarrassiBocquetBertinoEvensen2018,EvensenVossepoelvanLeeuwen2022} for just a few examples).

To describe our algorithm for parameter recovery, we shall denote the observable state of the velocity by $I_h(\mathbf{u})$, where $I_h$ is an (autonomous) bounded linear projection operator whose output is a suitable interpolation of the observed data into the phase space of the system \eqref{eq:NS_true}, which consists only of divergence-free vector fields. In general, $I_h$ is often referred to as an \textit{interpolant observable operator}. In the context of incompressible flows considered here, $I_h$ is understood to perform an interpolation of the observed data, then orthogonally projects the result onto solenoidal vector fields. Hence, $I_h^2=I_h$, $I_h\nabla r=0$, and $\nabla\cdotp I_h\mathbf{v}=0$, for any sufficiently smooth scalar field $r$ and vector field $\mathbf{v}$. Here, $h>0$ quantifies the resolution of the observational field in such a way that $\lim_{h\rightarrow 0}I_h(\mathbf{u})=\mathbf{u}$. 
Specific examples of the manner of interpolation encoded in $I_h$ include Lagrangian interpolation of nodal values or local averages which are distributed uniformly across the domain at a mesh-size $h$, but also include large-scale filtering such as projection onto finitely many Galerkin modes up to a cut-off frequency $1/h$. 

We will assume that $I_h$ and its complementary projection $J_h=I-I_h$ satisfy the following boundedness properties:
    \begin{align}\label{eq:boundedness}
        \|I_h(\partial^\alpha\phi)\|_2^2\leq {c_m^2}{h^{-2m}}\|\phi\|_2^2,\quad \|J_h(\phi)\|_2^2\leq C_m^2h^{2m}\sum_{|\alpha|=k}\|\partial^\alpha\phi\|_2^2
    \end{align}
for any multi-index $|\alpha|=m$ with $m\geq0$, for some constants $c_m, C_m>0$, independent of $h$; such inequalities are satisfied in the case where $I_h$ is given by spectral projection. In this special case, $I_h$ also satisfies $I_h\partial^\alpha=\partial^\alpha I_h$, so that $I_h(\nabla^2J_h(\mathbf{w}))=0$. For the remainder of the manuscript, $I_h$ is therefore assumed to represent spectral projection, that is, projection onto Fourier modes $|\mathbf{k}|\leq h^{-1}$, for $\mathbf{k}\in(\kappa_0\mathbb{Z})^2$.

The unobserved scales of the flow are then dynamically approximated by directly inserting the observations, $I_h(\mathbf{u})$, into \eqref{eq:NS_true} as a feedback-control term, resulting in the system
\begin{equation}\label{eq:NS_nudge}
    \partial_t\mathbf{v} + \mathbf{v}\cdot\nabla\mathbf{v}=- \nabla q + \nu \nabla^2 \mathbf{v} + \mathbf{g} - \mu I_h(\mathbf{v})+\mu I_h(\mathbf{u}),\quad \nabla\cdotp\mathbf{v}=0,
\end{equation}
where $\mu$ denotes a tuning parameter, often referred to as the \textit{nudging coefficient}, and $\mathbf{g}$ is a divergence-free vector field chosen as a putative approximation to the unknown forcing $\mathbf{f}$. A fundamental property of the system is that in the absence of model error, i.e., $\mathbf{g}=\mathbf{f}$, then the approximating flow field, $\mathbf{v}$, asymptotically synchronizes with the true flow field, $\mathbf{u}$, provided that the true flow field is observed through sufficiently small scales, i.e., $h\ll1$, and $\mu$ is appropriately tuned \cite{azouani2014continuous}.

In the presence of model error, this fundamental property can be leveraged to develop an ansatz for $\mathbf{f}$ by enforcing relaxation of the state error, $\mathbf{w}=\mathbf{v}-\mathbf{u}$, on the observed scales. Indeed, the evolution of $\mathbf{w}$ on the observed scales is governed by
    \begin{align}\notag
        \partial_tI_h(\mathbf{w})
        =I_h\left(-\mathbf{v}\cdotp\nabla\mathbf{v}+\nu\nabla^2\mathbf{v}+\mathbf{g}-\partial_t\mathbf{u}\right) -\mu I_h(\mathbf{w}).\notag
    \end{align}
Assuming that $\mathbf{g}$ can be chosen instantaneously such that
    \begin{align}\label{eq:ansatz}
        \mathbf{g}=\partial_tI_h(\mathbf{u})+I_h\left(\mathbf{v}\cdotp\nabla\mathbf{v}-\nu\nabla^2I_h\mathbf{u}-\nu\nabla^2J_h\mathbf{v}\right),
    \end{align}
where $J_h=I-I_h$ denotes the complementary projection, it follows that $\partial_tI_h(\mathbf{w})+\mu I_h(\mathbf{w})-\nu I_h\nabla^2J_h(\mathbf{w})=0$. Note that this choice critically relies on replacing the nudged velocity field $\mathbf{v}$ in the Laplacian term ($\nabla^2$) directly with the observed data $I_h\mathbf{u}$ combined with the unobserved data from the nudged solution $J_h\mathbf{v}$.  This is in direct contrast to simply identifying the forcing update as a function of $\mathbf{v}$.  This `direct replacement' strategy employed here is necessary for the rigorous convergence of the forcing to the true value, but as noted below does not appear to be necessary in practice.

By the orthogonality of $I_h(\mathbf{w})$ and $J_h(\mathbf{w})$, it follows that the energy balance at observational scales one step forward in time satisfies
    \begin{align}\label{eq:converge:low:modes}
        \|I_h(\mathbf{w})(t+\Delta t)\|_2=e^{-\mu\Delta t}\|I_h(\mathbf{w}(t))\|_2.
    \end{align}
In particular, exponential relaxation of $I_h(\mathbf{w})$ is enforced. The utility of this choice is readily seen; setting $\mathbf{h}=\mathbf{g}-\mathbf{f}$ we find that
    \begin{align}\label{eq:error:formal}
        I_h(\mathbf{h})
        &=I_h\left[J_h(\mathbf{w})\cdotp\nabla J_h(\mathbf{w})+J_h(\mathbf{w})\cdotp\nabla\mathbf{u}+\mathbf{u}\cdotp\nabla J_h(\mathbf{w})\right]\notag
        \\
        &\quad-\nu I_h(\nabla^2J_h\mathbf{w})+\mathcal{O}(I_h(\mathbf{w})).
    \end{align}
Under the assumption that $\mu$ is sufficiently large relative to the observational density, $h$, and time-step, $\Delta t$, the resulting state error one step forward, $\mathbf{w}(t+\Delta t)$ will satisfy the following estimates (see the Appendix for details):
    \begin{align}\label{eq:error:sync}
        \|\mathbf{w}(t+\Delta t)\|_2\leq \mathcal{O} \left(\frac{\|\mathbf{h}(t)\|_2}{\mu}\right),\quad \|\nabla\mathbf{w}(t+\Delta t)\|_2\leq \mathcal{O} \left(\frac{\|\mathbf{h}(t)\|_2}{\sqrt{\mu\nu}}\right).
    \end{align}

Further assuming that $I_h\mathbf{f}=\mathbf{f}$ (the forcing function lives in the observation space even though it is unobservable itself) it follows that $I_h\mathbf{h}=\mathbf{h}$. 
Then the model error in \eqref{eq:error:sync} is essentially controlled by
    \begin{align}
       \|I_h\mathbf{h}(t+\Delta t)\|_2&
       \leq \mathcal{O}(\|I_h(\mathbf{w}(t+\Delta t))\|_2)+\mathcal{O}\left(\frac{\|\mathbf{h}(t)\|_2}{\sqrt{\mu\nu}}\right)\notag.
    \end{align}
Hence, upon invoking \eqref{eq:converge:low:modes} for a sufficiently large time step $\Delta t$ and the fact that $I_h(\mathbf{h})=\mathbf{h}$, one may then deduce that for $\mu$ chosen appropriately large
    \begin{align}\label{eq:heuristic}
        \|\mathbf{h}(t+\Delta t)\|_2\leq\frac{1}2\|\mathbf{h}(t)\|_2.
    \end{align}
Owing to \eqref{eq:converge:low:modes}, we see that \eqref{eq:heuristic} implies exponential convergence of $\mathbf{h}$ to zero upon further iteration. 

\section{Definition of the algorithm}

The discussion above lends itself to an implementable algorithm:
At stage $1$, the algorithm is initialized with a pair $(\mathbf{v}_0^{-1}, \mathbf{g}^0)$ (the superscripts are indices, not exponentiation) that prescribes an initial velocity and a force for \eqref{eq:NS_nudge}. Integration of \eqref{eq:NS_nudge} forward-in-time produces $\mathbf{v}^0(t)=\mathbf{v}^0(t;\mathbf{v}_0^{-1},\mathbf{g}^0)$, for all times $t\in I_0=[0,\infty)$. After a transient period that allows the state error, $\mathbf{w}^0=\mathbf{v}^0-\mathbf{u}$ to establish a balance with the model error, $\Delta \mathbf{g}^0=\mathbf{g}^0-\mathbf{f}$, a new estimate of the forcing function modified from \eqref{eq:ansatz} is calculated at time $t_1\gg 0$:
    \begin{align}\label{eq:ansatz:mod:stage1}
        \mathbf{g}^1
        &=\partial_tI_h(\mathbf{u})+I_h\left(\mathbf{v}^0\cdotp\nabla\mathbf{v}^0-\nu\nabla^2(I_h\mathbf{u}+J_h\mathbf{v}^0)\right).
    \end{align}
This yields a new approximation, $\mathbf{g}^1$, to the force. Notably,  $\mathbf{g}^1=I_h(\mathbf{g}^1)$. This concludes the initial cycle of the algorithm. The process is then repeated: suppose that $\mathbf{g}^\ell$ have been produced in this way at stages $\ell=1,\dots, k$ at times $t_\ell\gg t_{\ell-1}$, respectively, and that $\mathbf{g}^\ell=I_h\mathbf{g}^\ell$. At stage $k$, \eqref{eq:NS_nudge} is re-initialized at time $t=t_k$, with the pair $(\mathbf{v}^{k-1}(t_k),\mathbf{g}^k)$; this produces an approximating velocity, $\mathbf{v}^{k}(t)=\mathbf{v}^k(t;\mathbf{v}^{k-1}(t_k),\mathbf{g}^k)$, over the time interval $I_k=[t_k,\infty)$. After a sufficiently long interval of time for $\mathbf{w}^k=\mathbf{v}^k-\mathbf{u}$ to achieve balance with the model error, $\Delta\mathbf{g}^k=\mathbf{g}^k-\mathbf{f}$, the forcing function estimate is updated according to
        \begin{align}\label{eq:ansatz:mod:stagekp1}
        &\mathbf{g}^{k+1}=\partial_tI_h(\mathbf{u})+I_h\left(\mathbf{v}^k\cdotp\nabla\mathbf{v}^k-\nu\nabla^2(I_h\mathbf{u}+J_h\mathbf{v}^k)\right),
    \end{align}
at time $t_{k+1}\gg t_{k}$ to produce $\mathbf{g}^{k+1}$; again $\mathbf{g}^{k+1}=I_h\mathbf{g}^{k+1}$. In this way a sequence of approximating forces is generated, $\mathbf{g}^1,\mathbf{g}^2,\mathbf{g}^3,\dots$, as observations are assimilated. 

\begin{Rmk}\label{rmk:algorithms}
As mentioned in the introduction, the derivation of the algorithm presented in the current article allows one to obtain many other algorithms. In particular, the algorithm studied in \cite{martinez2022force} is obtained by replacing the $\mathbf{v}$ with $I_h\mathbf{u}+(I-I_h)\mathbf{v}$. All other variations of the algorithm may be obtained by applying this substitution in some, rather than all of the terms where $\mathbf{v}$ appears in \eqref{eq:ansatz:mod:stagekp1}. In this way, the point of view introduced in \cite{pachev2022concurrent} and subsequently adopted here allows for a more general perspective. 

One may notice in \eqref{eq:ansatz:mod:stagekp1} that $I_h\mathbf{u}$ is being used directly in the viscous term. At the moment, the analysis performed below is unable to replace this specific term $I_h\mathbf{v}$. We believe this to be a technical point, however, since the numerical tests performed below suggest that one should nevertheless achieve convergence without this `direct replacement' employed. Nevertheless, establishing convergence of the version of the algorithm corresponding to \eqref{eq:ansatz} remains an open problem.
\end{Rmk}

\begin{Rmk}\label{rmk:shear}
In the setting presented above, we have assumed that the velocity field is mean-free. This assumption is typically justified by recalling the Galilean invariance of the Navier-Stokes equations that allows one to shift the reference frame of the fluid in order to equivalently view it with respect to one that is mean-free. Coupled with the observation that the mean velocity 
is preserved under the free-evolution of the Navier-Stokes equation, that is, in the absence of external force, it then suffices to consider the scenario where the initial fluid velocity is mean-free. In the presence of an external force, the mean-free assumption remains valid provided that the external force is also assumed to be mean-free.  This assumption is not essential for the analysis provided here, but does make the rigorous proof much simpler to establish.

Our setting therefore implicitly assumes that 1) the force is restricted to the observational subspace, i.e., $I_h\mathbf{f}=\mathbf{f}$ as is apparently necessary as indicated in FIG \ref{conv_plots_H1}, 2) that the force is periodic in space, and 3) that the mean-force is zero, i.e., $\bar{\mathbf{f}}=|\Omega|^{-1}\int_\Omega \mathbf{f}(x)dx=0$. The third assumption is not more restrictive than the assumption that one has access to all large-scale motions of a fluid down to a particular length scale $h>0$. Indeed, since $I_h$ is assumed to be the projection onto the subspace of Fourier modes up to wave-number $|k|\leq h^{-1}$, where $k\in (\kappa_0\mathbb{Z})^2$, and observations are made continuously in time, if $\mathbf{f}$ happens to possess a non-zero mean, then one easily sees that $\frac{d}{dt}I_{h>1}u=I_{h>1}\mathbf{f}=\hat{\mathbf{f}}(0)$. In particular, in the setting we study, it is implicitly assumed that $\hat{\mathbf{f}}(0)$ is  accessible to the user from direct observations.

One particular scenario of interest in which the fluid possesses a nonzero mean is that of shear flows with nonzero mean profile. Due to the remarks above, this scenario is indeed within the purview of the setting studied here. However, it is not interesting in the absence of boundaries, which is the main setting of this article. The difficult problem of rigorously studying the effects of boundaries, as they arise for instance from no-slip boundary conditions, is relegated to future work. We simply point out that the choice of considering the setting of 2D isotropic turbulence represents an important {first step} towards understanding the problem of inferring unknown external forces from low-mode observations in turbulent flows. 
\end{Rmk}

\section{Rigorous Convergence Analysis}\label{sec:analysis}

The precise version of \eqref{eq:error:sync} asserts that the stage $k$ state error satisfies  
    \begin{align}\label{eq:error:sync:true}
        \|\mathbf{w}^k(t)\|_2\leq \mathcal{O}\left(\frac{\|\Delta\mathbf{g}^{k}\|_2}{\mu}\right),\quad
        \|\nabla\mathbf{w}^k(t)\|_2\leq \mathcal{O}\left(\frac{\|\Delta\mathbf{g}^{k}\|_2}{\sqrt{\mu\nu}}\right),
    \end{align}
for all $t\geq t_{k+1}$, where $t_{k+1}\gg t_k$, provided that $\mu h^2\leq \nu$ and $\mu$ is chosen sufficiently large, depending only on the maximum magnitude of the energy, enstrophy, and palenstrophy of $\mathbf{u}$; a proof of the first inequality in \eqref{eq:error:sync:true} is provided in Section \ref{sec:appendix}, while the proof of the second can be found in \cite{martinez2022force}.

Now, from \eqref{eq:NS_true}, \eqref{eq:NS_nudge}, and \eqref{eq:ansatz}, the model error at stage $k+1$ can be expanded as:
    \begin{align}\label{eq:error:rep}
        \Delta\mathbf{g}^{k+1}
        &=\left[I_h\left(\mathbf{w}^{k}\cdotp\nabla\mathbf{w}^{k}\right)+I_h\left(\mathbf{u}\cdotp\nabla\mathbf{w}^{k}\right)+I_h\left(\mathbf{w}^{k}\cdotp\nabla\mathbf{u}\right)\right]_{t=t_{k+1}}.
    \end{align}
Note that the fact that $\mathbf{f}$ is confined to the subspace spanned by the observations, that is, $\mathbf{f}=I_h\mathbf{f}$, has been applied. It is then clear from \eqref{eq:error:rep} that $I_h\Delta \mathbf{g}^{k+1}=\mathbf{g}^{k+1}$.
To assess the size of $\Delta\mathbf{g}^{k+1}$, observe that integration by parts and the properties $I_h^2=I_h$, $I_h\mathbf{g}^k=\mathbf{g}^k$, for all $k$, yield
    \begin{align}
        \langle I_h\left(\mathbf{w}^{k}\cdotp\nabla\mathbf{w}^{k}\right),\Delta\mathbf{g}^{k+1}\rangle&=-\langle \mathbf{w}^{k}\cdotp\nabla \Delta\mathbf{g}^{k+1},\mathbf{w}^{k}\rangle, \notag\\
        \langle I_h(\mathbf{u}\cdotp\nabla\mathbf{w}^k),\Delta\mathbf{g}^{k+1}\rangle&=-\langle \mathbf{u}\cdotp\nabla \Delta\mathbf{g}^{k+1},\mathbf{w}^k\rangle,\notag\\
        \langle I_h(\mathbf{w}^k\cdotp\nabla\mathbf{u}),\Delta\mathbf{g}^{k+1}\rangle&=-\langle \mathbf{w}^k\cdotp\nabla\Delta\mathbf{g}^{k+1},\mathbf{u}\rangle\notag,
    \end{align}
where brackets denote the $L^2$--inner product over $\Omega$. Now, upon taking the $L^2$--inner product of \eqref{eq:error:rep} with $\Delta\mathbf{g}^{k+1}$, integrating by parts, invoking the fact that $\Delta\mathbf{g}^{k+1}$ is solenoidal, then applying the Cauchy-Schwarz inequality, H\"older's inequality (see the Appendix), and \eqref{eq:boundedness}, yields
    \begin{align}\notag
    \|\Delta\mathbf{g}^{k+1}\|_2^2&\leq \left(\|\mathbf{w}^{k}\|_{4}^2
    +2\|\mathbf{u}\|_{\infty}\|\mathbf{w}^{k}\|_2\right)\|\nabla\Delta\mathbf{g}^{k+1}\|_2.
    \end{align}
Let $R_k$ denote the supremum in time of $\||\nabla|^k\mathbf{u}(t)\|_2$. By interpolation, the Cauchy-Schwarz inequality, and \eqref{eq:boundedness} it follows that
    \begin{align}
        \|\Delta\mathbf{g}^{k+1}\|_2\leq  \frac{c_0}{h}\left(\|\nabla\mathbf{w}^k\|_2\|\mathbf{w}^k\|_2+\sqrt{R_2R_0}\|\mathbf{w}^k\|_2\right),\notag 
    \end{align}
for some universal non-dimensional constant $c_0>0$ that depends on $c_1, c_2$ from \eqref{eq:boundedness} and the constants of interpolation; we refer to the Appendix for additional details. Finally, \eqref{eq:error:sync:true} implies
    \begin{align}\label{eq:pre:convergence}
     \|\Delta\mathbf{g}^{k+1}\|_2\leq c_0\frac{\sqrt{R_2R_0}}{\mu h}\left(\frac{\|\Delta\mathbf{g}^{k}\|_2}{\sqrt{\mu\nu R_2R_0}}+1\right)\|\Delta\mathbf{g}^{k}\|_2,
    \end{align}
where $c_0$ now denotes a larger, but still non-dimensional, constant. Conditions ensuring the convergence of $\mathbf{g}^k$ to the true force, $\mathbf{f}$, may then be given by
    \begin{align}\label{eq:conditions}
    2c_0(R_2R_0)^{1/2}h\leq \mu h^2\leq\nu,\quad \mu\nu \geq \frac{1}{2}\frac{\|\Delta\mathbf{g}^0\|_2}{\sqrt{R_2R_0}}.
    \end{align}
Indeed, under \eqref{eq:conditions}, one may then deduce from \eqref{eq:pre:convergence} that
    \begin{align}
          \|\Delta\mathbf{g}^{k+1}\|_2\leq \frac{1}2\|\Delta\mathbf{g}^{k}\|_2,\quad\text{for all}\ k\geq0.
    \end{align}
Observe that \eqref{eq:conditions} constitutes a non-trivial set of pairs $(\mu,h)$ whenever the observational density, $h^{-1}$, is sufficiently large in relation to the viscosity, energy ($R_0$), and palenstrophy ($R_2$), of the flow, and the nudging parameter, $\mu$, is tuned appropriately large. Note that the choice of $\mu$ ultimately depends on the true forcing, $\mathbf{f}$, through the values of $R_0,~R_2$ and the initial error. In practice, this is often approximately known, for instance, in terms of the Reynolds number of the flow; we refer the reader to the Appendix for an additional comment on this point.

\section{Simulation Results}\label{sec:simulations}
To demonstrate the utility of this algorithm in practice, we computationally recover the forcing function for 2D forced incompressible Navier-Stokes.
\begin{itemize}
    \item The simulation was carried out in Matlab R2021a using a pseudo-spectral method with explicit Euler time-stepping, respecting the CFL constraint and 2/3-dealiasing rule, and the mean-free condition enforced.  The spatial resolution was $2048^2$  (giving roughly $(2048\cdot2/3)^2-1$ degrees of freedom) on the domain $[-\pi,\pi)^2 $ with time-step 
% $\delta t =3.125\times 10^{-4}$.  
$\Delta t =0.0025$. 
The reference $\mathbf{u}$ system was simulated at the stream-function level using the Basdevant formula \cite{basdevant1983technical} to compute the nonlinearity efficiently.   Namely, the following equivalent form of the 2D Navier-Stokes equations were used:
\[\psi_t = \nabla^{-2}[(\partial_x^2-\partial_y^2)(u_1u_2) + \partial_{xy}(u_2^2 - u_1^2)] + f^\psi\]
where $\nabla^\perp:=\binom{-\partial_y}{\partial_x}$, $\mathbf{u}=\binom{u_1}{u_2}= \nabla^\perp\psi$, and $\mathbf{f}=\nabla^\perp f^\psi$.  The equation for the data assimilation $\mathbf{v}$ system was handled similarly.  
%https://www.overleaf.com/project/62dec3e03013adf33497c1fb
\item The unobserved forcing function $f^\psi$ was determined by randomly (normal, $N(0,1)$, distribution-seeded in Matlab with \texttt{rng(0,\textquotesingle twister\textquotesingle)}) picking amplitudes for wave-modes in Fourier space on an annulus with inner radius 16 and outer radius 64 (see FIG. \ref{fig:a} for an illustration of the forcing function). Specifically, the forcing recovery algorithm was tasked to recover 11,672 (real-valued) unknowns of the forcing.
\item To ensure that the simulation is well-resolved, and also understand the behavior at various wave lengths, we plot the (weighted) spectrum of various functions.  Namely, given a function $\rho$ where $\rho(\mathbf{x}) = \sum_{\mathbf{n}\in\mathbb{Z}^2} \widehat{\rho}_\mathbf{n}e^{i\mathbf{n}\cdot\mathbf{x}}$, 
we denote its weighted spectrum by
\[E_s[\rho](k):=
(k+1)^{s}\Big(\sum_{\substack{\mathbf{n}\in\mathbb{Z}^2\\k\leq|\mathbf{n}|<k+1}} 
|\widehat{\rho}_\mathbf{n}|^2\Big)^{1/2}
\] 
where $s\in\{0,1,2\}$ is a weighting exponent for illustrative purposes, and $k\in\mathbb{N}$ is referred to as the ``wave number.''  
Roughly speaking, for $\mathbf{u}=\nabla^\perp\psi$, $E_1(\psi)$ corresponds to the kinetic energy spectrum (of $\mathbf{u}$) and $E_2(\psi)$ corresponds to the enstrophy spectrum. 
Due to the mean-free condition, $E_s[h](0)=0$; hence it is not plotted.
\item The viscosity was $\nu=10^{-4}$, and the Grashof number was $G=2.5\times 10^6$, leading to a solution with an energy spectrum resolved to machine precision ($\approx 2.22\times 10^{-16}$) before the dealiasing cut-off at $k=2*2048/3$.
\item Initial data for $\psi$ (i.e., for the $\mathbf{u}$ system) was given by a simulation spun up from initially-zero data and run with the same forcing until energy and enstrophy were roughly statistically stabilized; namely, time $t=210$ (see FIG. \ref{fig:a}).  That is, no major qualitative differences were observed in the spectrum for $t\geq210$.  It turns out that with these parameters, the initial $\psi$ is fairly small; namely, $\|\psi\|_{L^2}\approx 0.0414346$.  However, this is comparable with initial data in other studies with similar parameters, such as \cite{GeshoOlsonTiti2016,Olson_Titi_2008_TCFD}.

\begin{figure}[ht]
%\centering
\begin{subfigure}[b]{0.49\textwidth}
\includegraphics[width=1\textwidth,trim = 0mm 0mm 0mm 0mm, clip]{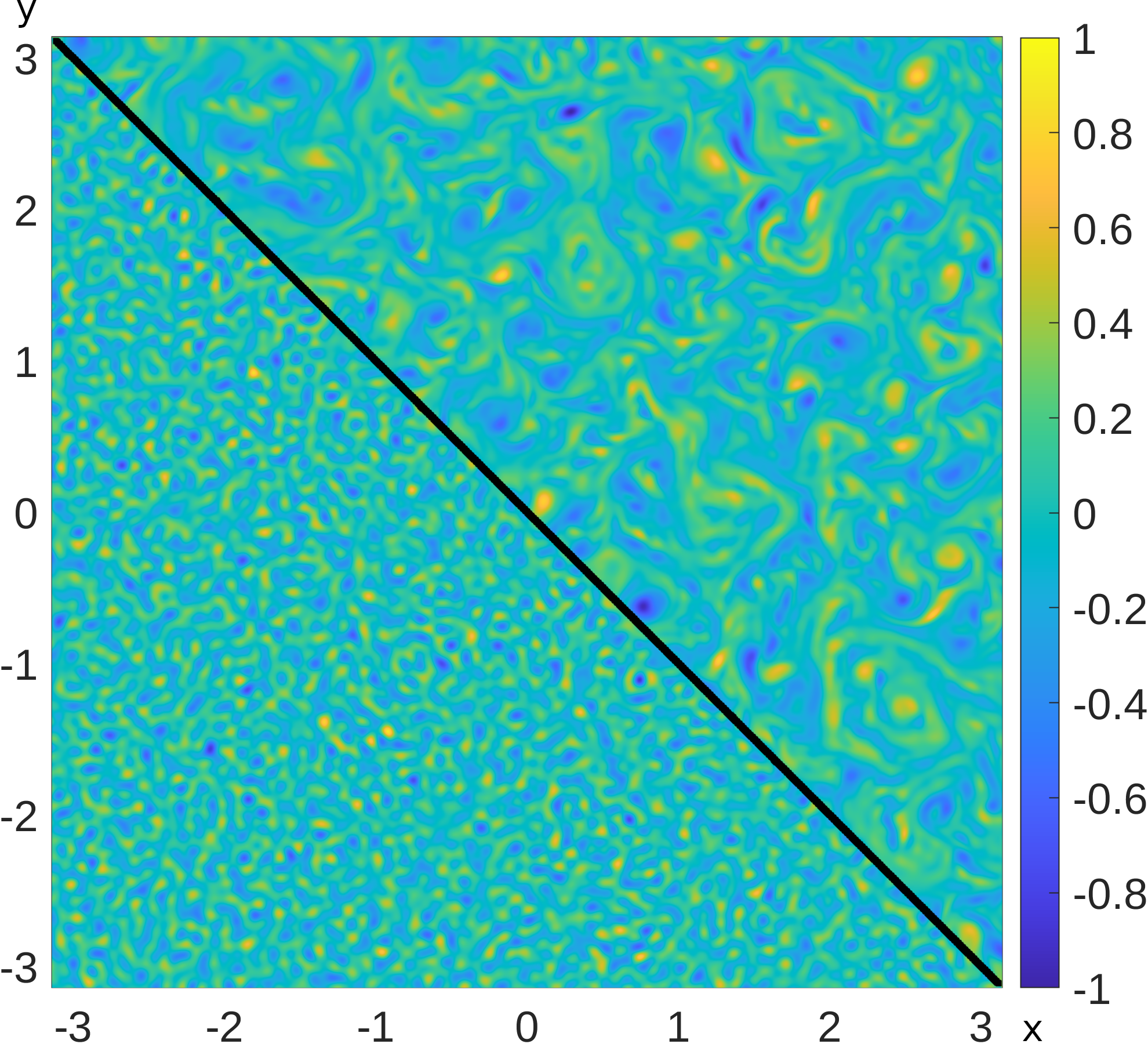}
\caption{\tabular[t]{@{}l@{}}Visualization of reference force and initial data,
\\
(bottom-left cut-away) $\nabla^2 f^\psi/\|\nabla^2 f^\psi\|_{L^\infty}$,
\\
(top-right cut-away) $\nabla^2 \psi_0/\|\nabla^2 \psi_0\|_{L^\infty}$.
\endtabular}
\end{subfigure}
\begin{subfigure}[b]{0.47\textwidth}
\includegraphics[width=1\textwidth,trim = 0mm 0mm 0mm 0mm, clip]{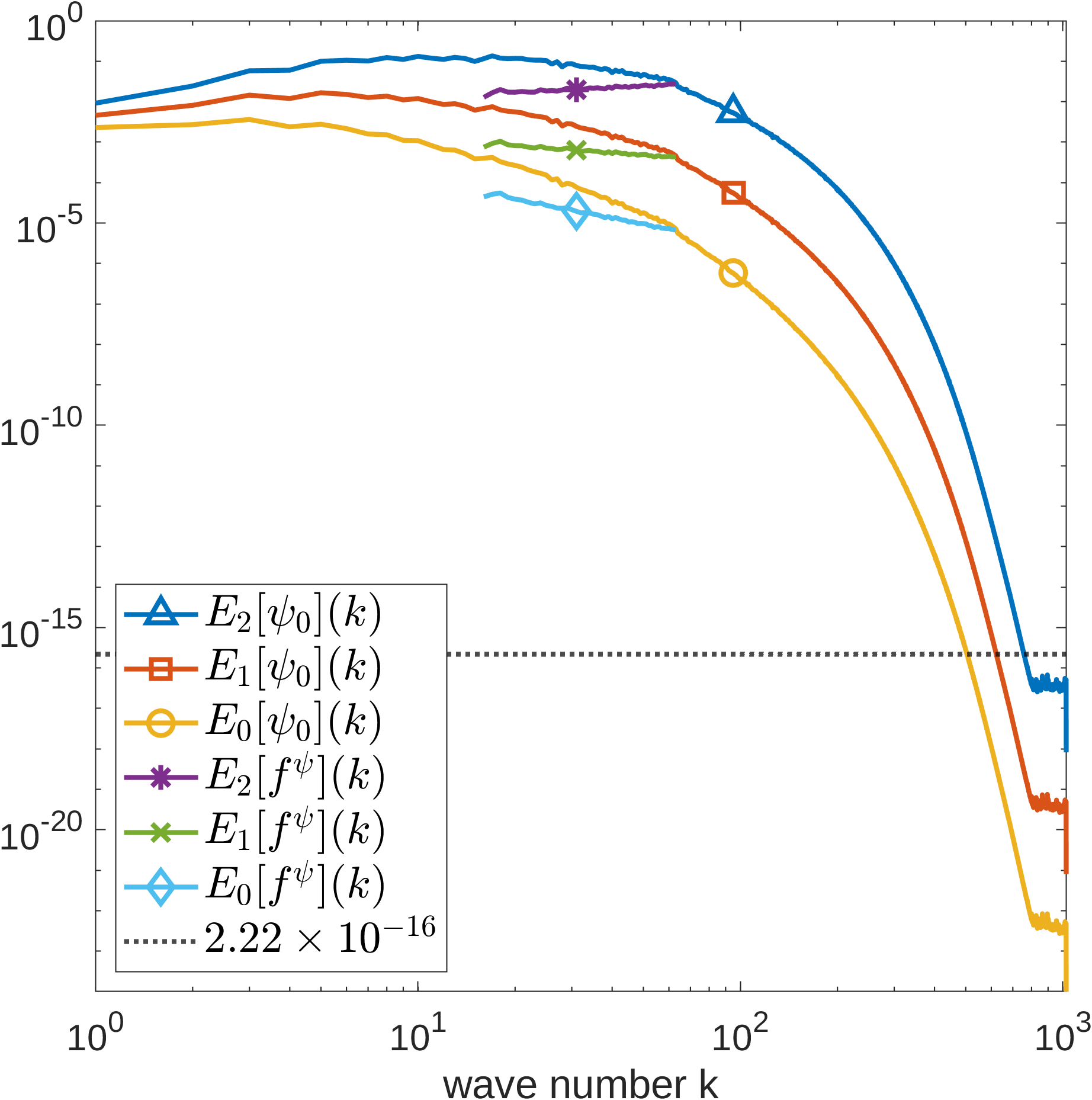}
\caption{Weighted spectra of force $f^\psi$ and initial data $\psi_0$.}
\end{subfigure}
\caption{\label{fig:a} Initial data and reference force.  In (a), the normalized initial vorticity $-\nabla^2 \psi_0/\|\nabla^2 \psi_0\|_{L^\infty}$ is shown rather than the initial stream function $\psi_0$, to highlight details.  Similarly, $-\nabla^2 f^\psi/\|\nabla^2 f^\psi\|_{L^\infty}$ is shown in (a) rather than $f^\psi$. For reference, we computed that $\|\nabla^2 \psi_0\|_{L^\infty}\approx2.8256$ and $\|\nabla^2 f^\psi\|_{L^\infty}\approx0.7443$.  In (b), the weighted spectra of the initial data and stream function are shown.}
\end{figure}
\item Initial data and forcing for the $\mathbf{v}$ system were identically zero.  

% \begin{figure}[h]
% %\centering
% \begin{subfigure}[b]{0.49\textwidth}
% \includegraphics[width=1\textwidth,trim = 0mm 0mm 0mm 0mm, clip]{halfhalf_Centered.png}
% \caption{Force $f^\psi$ 
% (bottom-left cut-away, left color bar range: $[-2.7,3.1]$) 
% and initial vorticity $\triangle \psi$ 
% (top-right cut-away, right color bar range: $[-0.73,0.64]$).  
% $xy$-axis: $[-\pi,\pi)\times[-\pi,\pi)$.}
% \end{subfigure}
% \begin{subfigure}[b]{0.49\textwidth}
% \includegraphics[width=1\textwidth,trim = 0mm 0mm 0mm 0mm, clip]{IC_force_spec.png}
% \caption{Weighted spectra of force $f^\psi$ and initial data $\psi$.}
% \end{subfigure}
% \caption{\label{fig:a} Initial data and forcing.  In figure (a), the  vorticity $\nabla^\perp\mathbf{u}=-\triangle \psi$ is shown rather than the stream function $\psi$, as it is easier to see details.}
% \end{figure}
% \item Initial data and forcing for the $\mathbf{v}$ system were identically zero.  

\item The interpolation operator, $I_h$, was taken to be a projection onto the ``observed'' Fourier modes of the stream function, that is, those on wave-modes $k$, $0<|\mathbf{k}|_\infty\leq 64$.  In particular,  only $\approx0.893\%$ of the unknowns in the solution were observed.  The nudging parameter was taken to be $\mu=1.9/\Delta t$.

\item The time-derivative that appears in the ansatz forcing function \eqref{eq:ansatz:mod:stagekp1} was computed via a forward-Euler finite difference approximation of $\partial_t I_h(\mathbf{v})$.

\item The nonlinearity of the nudged equation was computed the same way as in the solution of the `truth' system (using $\mathbf{v}$ instead of $\mathbf{u}$), and the same can be said for the forcing estimate in \eqref{eq:ansatz:mod:stagekp1}.

\item When we refer to the ``$L^2$ Error'' in figures or the text (say, the $L^2$ error between $\psi$ and $\psi_{DA}$), we specifically mean the following:
\begin{align}
    \|\psi - \psi_{DA}\|_{L^2}:=\left(\int_{[-\pi,\pi)^2}|\psi(\mathbf{x}) - \psi_{DA}(\mathbf{x})|^2\,d\mathbf{x}\right)^{1/2}
\end{align}
In particular, no normalization factor is used.  Of course, in simulations, this is computed discretely (via Parseval's identity), which we do via a Riemann sum.
\end{itemize}

\begin{figure}[ht]
\begin{subfigure}[b]{0.49\textwidth}
\includegraphics[width=1\textwidth,trim = 0mm 0mm 0mm 0mm, clip]{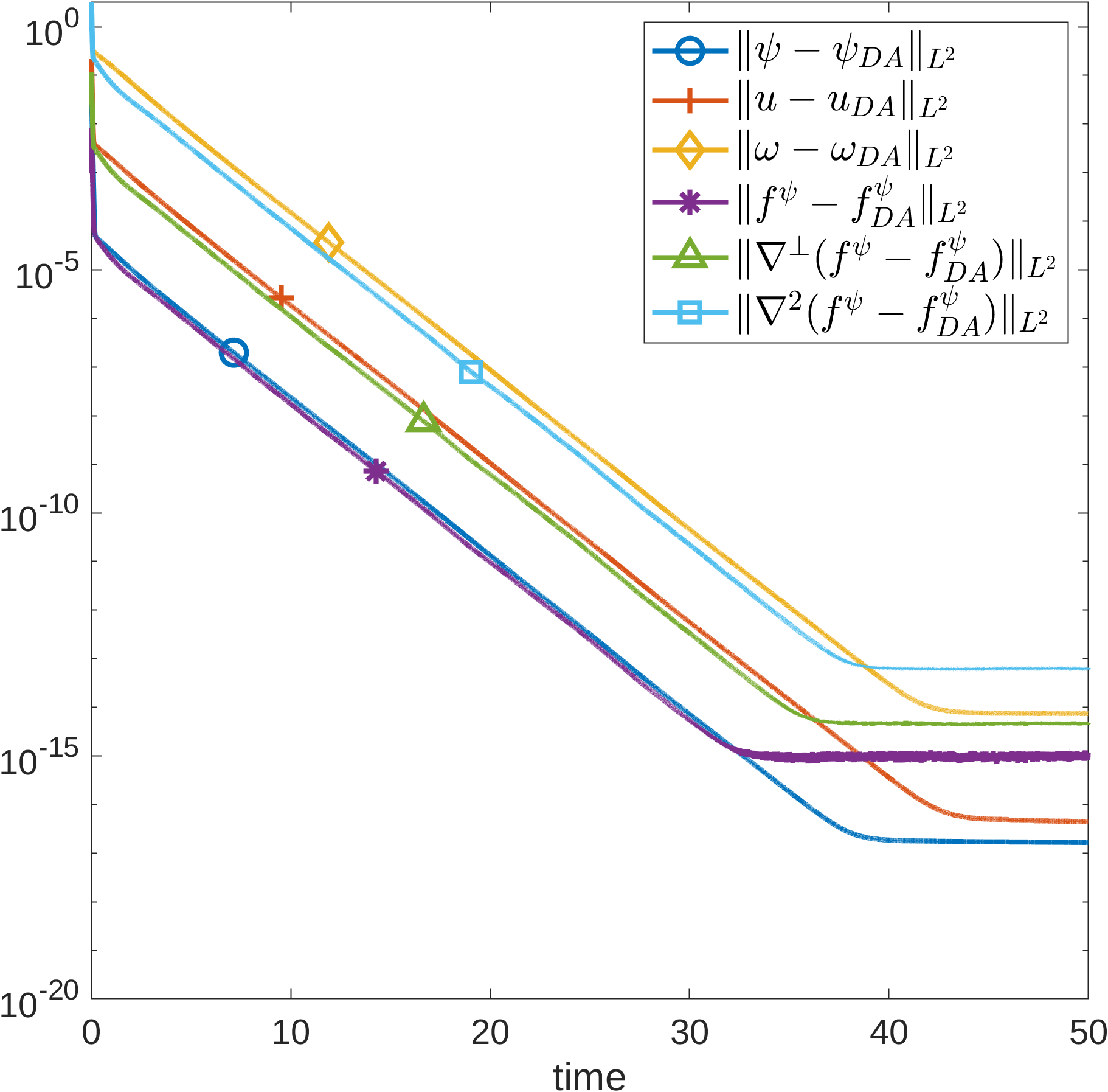 }
\caption{Observation of all forcing modes, i.e., $|\mathbf{k}|_\infty\leq64$.  Solution and forcing converge exponentially fast.}
\end{subfigure}
\begin{subfigure}[b]{0.49\textwidth}
\includegraphics[width=1\textwidth,trim = 0mm 0mm 0mm 0mm, clip]{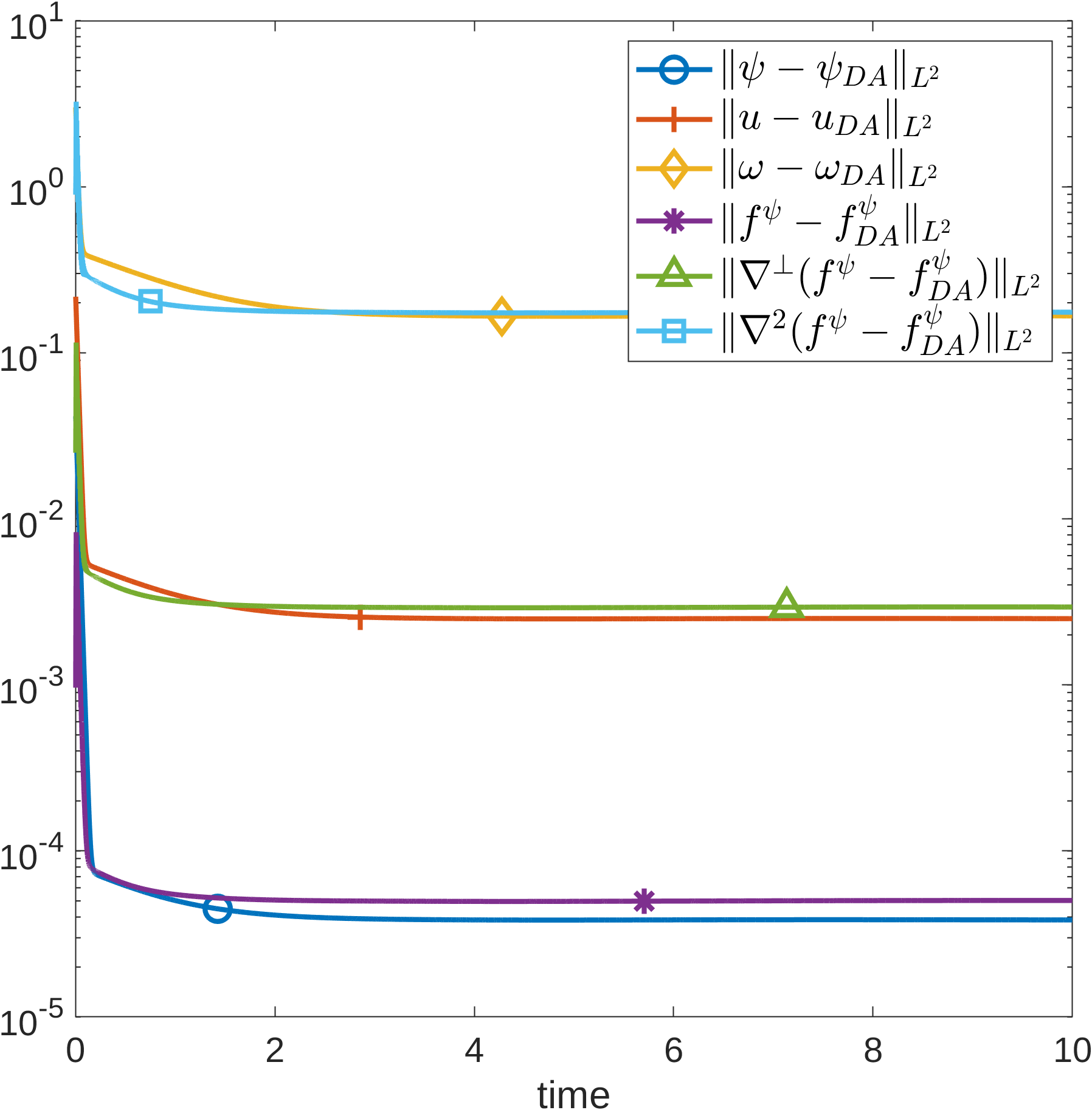}
\caption{Observation of only forcing modes $\mathbf{k}$ with $|\mathbf{k}|_\infty\leq60$.  Convergence is halted around error $\approx10^{-4}$.}
\end{subfigure}
\caption{\label{conv_plots_H1} Convergence plots for the exact Laplacian algorithm with force updated continuously (i.e., at each time step) in various norms.  Comparing (a) and (b), it appears that observing all modes contained in the force gives convergence to machine precision, while not observing only a small number of modes leads to lack of convergence.}
\end{figure}

Convergence of the algorithm relies on the the number of observed modes in a discontinuous fashion as displayed in FIG \ref{conv_plots_H1}.  Note that any additional modes exceeding the number contained in the forcing function do not noticeably affect the convergence rate or the final error level (which is near machine epsilon anyway).  However, not observing any of the modes in the flow field that correspond to the unknown modes in the force lead to a lack of complete convergence either of the forcing function or the flow field itself.  This indicates that the restriction of observing the modes (scales) in the flow field that correspond to the relevant scales in the forcing function itself, is (at least numerically) quite strict.

\begin{figure}[ht]
\includegraphics[width=1.0\linewidth,trim = 0mm 0mm 0mm 0mm, clip]{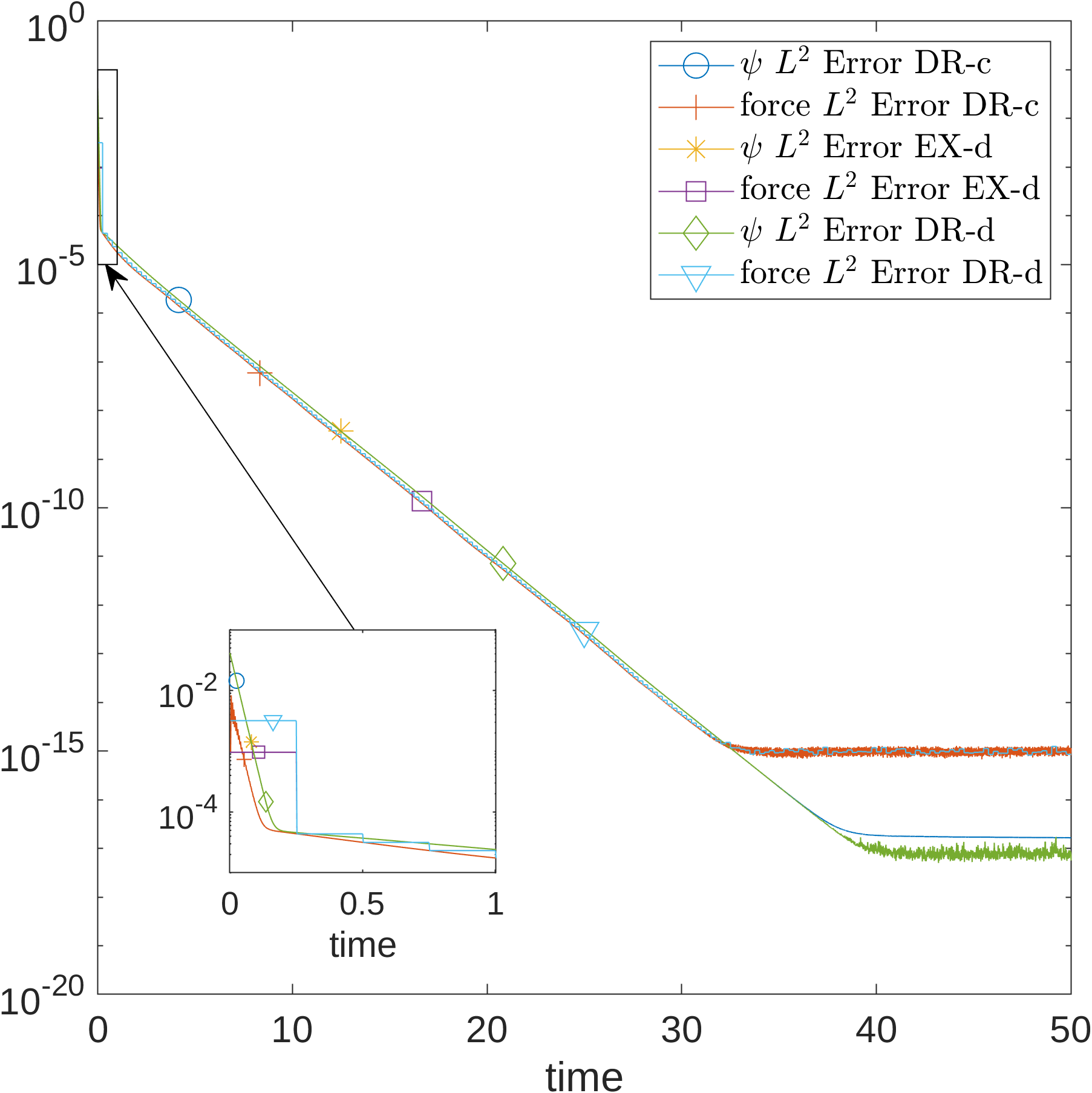}
\caption{\label{fig:b} (log-linear plot) $L^2$ errors vs. time.  ``DR'' refers to the direct replacement algorithm, while ``EX'' refers to the exact Laplacian algorithm.  ``-c'' indicates that the force was updated continuously (i.e., at each time step), while ``-d'' indicates discrete force updates, every $0.25$ time units.}
\end{figure}

The forcing estimation was implemented in two different ways:
\begin{enumerate}
    \item As outlined here, a direct replacement strategy was used for the Laplacian term, i.e. \eqref{eq:ansatz:mod:stagekp1} was used.  This is the method justified by the rigorous analysis.
    \item Rather than replace the Laplacian term with a combination of the low modes from the observed truth $I_h\mathbf{u}$ and the high modes of the nudged system $J_h\mathbf{v}$, we also simply used the low modes of the nudged system, i.e. replacing \eqref{eq:ansatz:mod:stagekp1} with
    \begin{equation}\label{eq:ansatz:exact}
        \mathbf{g}^{k+1} = \partial_t I_h(\mathbf{u}) + I_h(\mathbf{v}^k\cdot \nabla \mathbf{v}^k - \nu \nabla^2  \mathbf{v}^k).
    \end{equation}
\end{enumerate}
Although only the first case is rigorously justified, we found no qualitative computational difference between these two schemes, i.e. convergence of both the state and approximated forcing function were nearly identical between the two approaches.  This is demonstrated in FIG \ref{fig:b}.  In this Figure, the direct replacement (DR) simulations are those where the forcing update is given by \eqref{eq:ansatz:mod:stagekp1} whereas the exact Laplacian refers to \eqref{eq:ansatz:exact}.  Note that there is very minimal difference between the convergence rates in either case.

\begin{remark}
% First draft:
%The constant $\mu>0$ is a user-tunable input parameter, and one just wants it to be as large as possible before either (a) it violates numerical stability (which is why we choose $\mu < 2/\Delta t$, or (b) it starts to destabilize the high modes (this does not happen in our case, since $I_h$ is just Fourier truncation, and hence the only influence is on the low modes, but in the case when $I_h$ is, e.g., piecewise-linear interpolation, high modes can be destabilized.  This “spill-over” effect is discussed at length in Appendix 7.1 of [Carlson-Larios-Titi].)
% Second draft:
%The parameter $\mu>0$ serves as a user-adjustable input within our framework, optimized by increasing its value up to a threshold defined by two critical constraints. Firstly, to preserve numerical stability for the explicit Euler step, we ensure $\mu<2/\Delta t$. Secondly, while our specific application involves Fourier truncation, which primarily affects low modes without destabilizing higher modes, it is worth noting that alternative implementations, such as piecewise-linear interpolation, could introduce instability in these higher modes. This phenomenon, sometimes called the ``spill-over'' effect, is discussed at length in Appendix 7.1 of \cite{}, highlighting the nuanced balance required in parameter tuning to maintain the integrity of the system's high-frequency components.
The parameter $\mu>0$ serves as a user-adjustable input, optimized by increasing its value (resulting in better convergence rates) up to a threshold determined by two forms of stability constraints.  %Since it is non-physical and a user-input, %it is not particularly interesting to consider different values of $\mu$; 
In practice one simply increases it in preliminary simulations until just before error rates go back up again, and this value (scaling appropriately with respect to $\Delta t$) should work for more involved simulations with similar numerical discretization and parameter regimes.  The constraints to keep in mind are: (i) to ensure numerical stability (e.g., for the explicit Euler scheme) for the explicit Euler step, one should ensure the basic CFL-type constraint that $\mu<2/\Delta t$, and (ii) sufficiently large $\mu$ may, in some cases, destabilize the small scales.  Constraint (ii) this does not arise in our case, because $I_h$ is simple Fourier truncation, and hence the only influence is on the large scales, and hence we only require $\mu<2/\Delta t$. This ultimately informed our choice of $\mu=1.9/\Delta t$.  Slightly larger values of $\mu$ are possible, but only yield marginally better convergence rates.  Choosing $\mu\geq2/\Delta t$ yields unstable simulations as expected.)  However, in the case when $I_h$ is, e.g., piecewise-linear interpolation, high modes can be destabilized.  This ``spill-over'' effect is discussed at length in Appendix 7.1 of \cite{Carlson_Larios_Titi_2023_nlDA}.  While it might be interesting for theoretical reasons to estimate a range of $\mu$-values which lead to spill-over for a given choice of $\nu$, $h$, etc., as mentioned, in practice it is fairly straight-forward to find this range in numerical experiments, and in any case, it would not affect the results of the present work due to our choice of $I_h$.
\end{remark}

\begin{figure}[ht]
\begin{subfigure}[b]{0.49\textwidth}
\includegraphics[width=1.0\linewidth,trim = 0mm 0mm 0mm 0mm, clip]{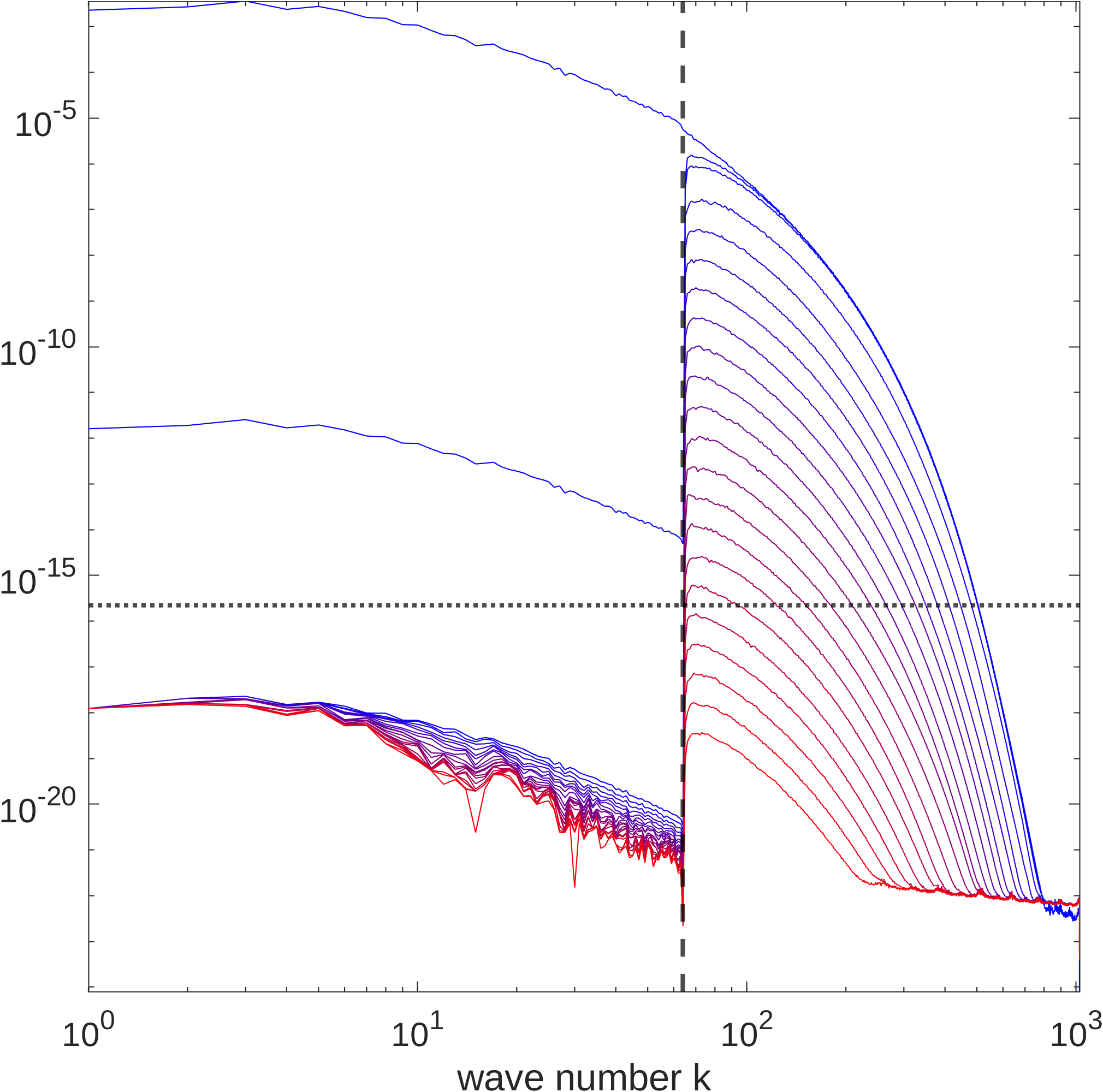}
\caption{Spectral error in stream function: $E_0[\psi - \psi_{DA}](k)$ vs. $k$.}
\end{subfigure}
\begin{subfigure}[b]{0.49\textwidth}
\includegraphics[width=1.0\linewidth,trim = 0mm 0mm 0mm 0mm, clip]{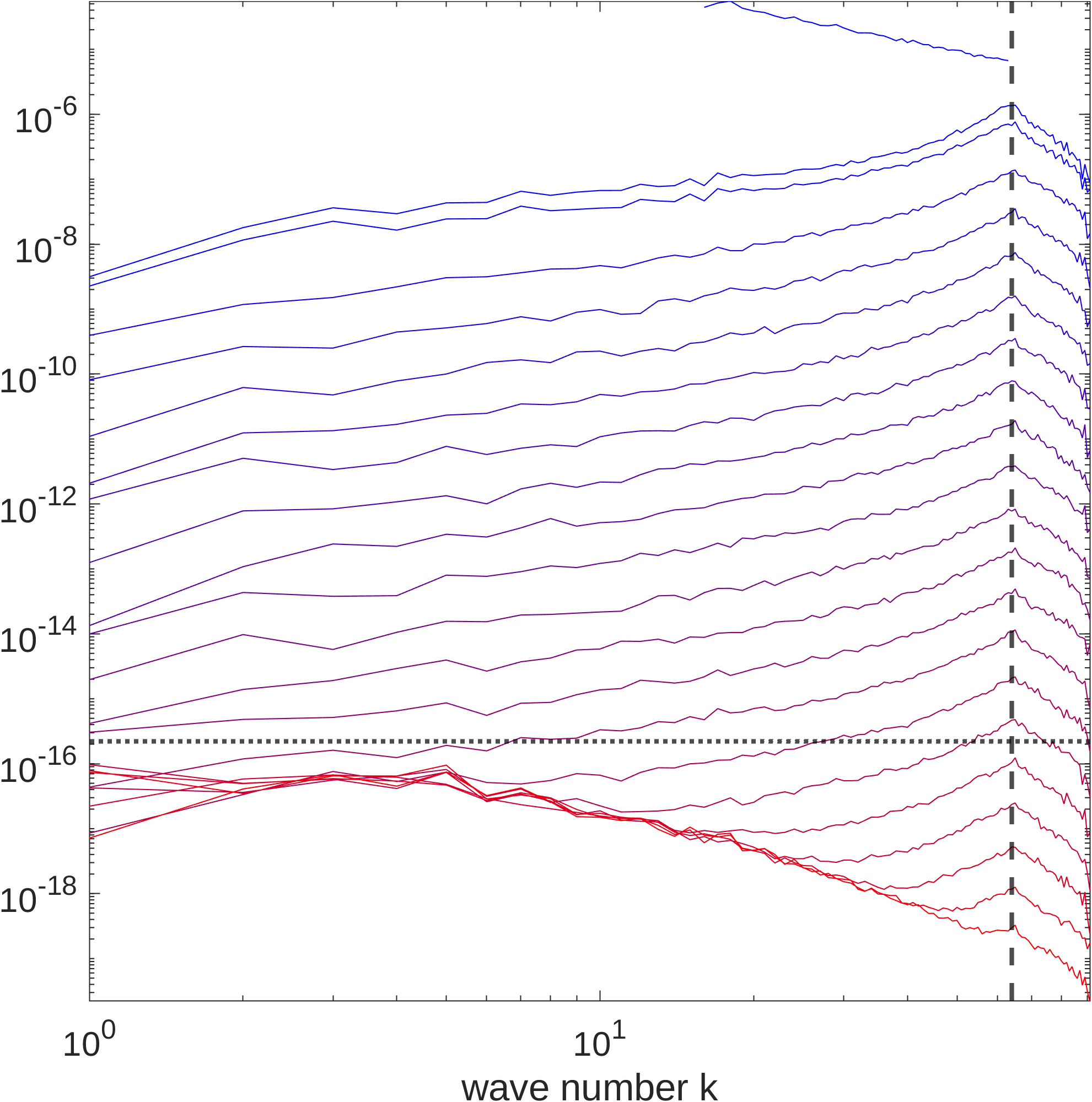}
\caption{Spectral error in forcing: $E_0[f^\psi - f^\psi_{DA}](k)$ vs. $k$.}
\end{subfigure}
\begin{subfigure}[b]{0.49\textwidth}
\includegraphics[width=1.0\linewidth,trim = 0mm 0mm 0mm 0mm, clip]{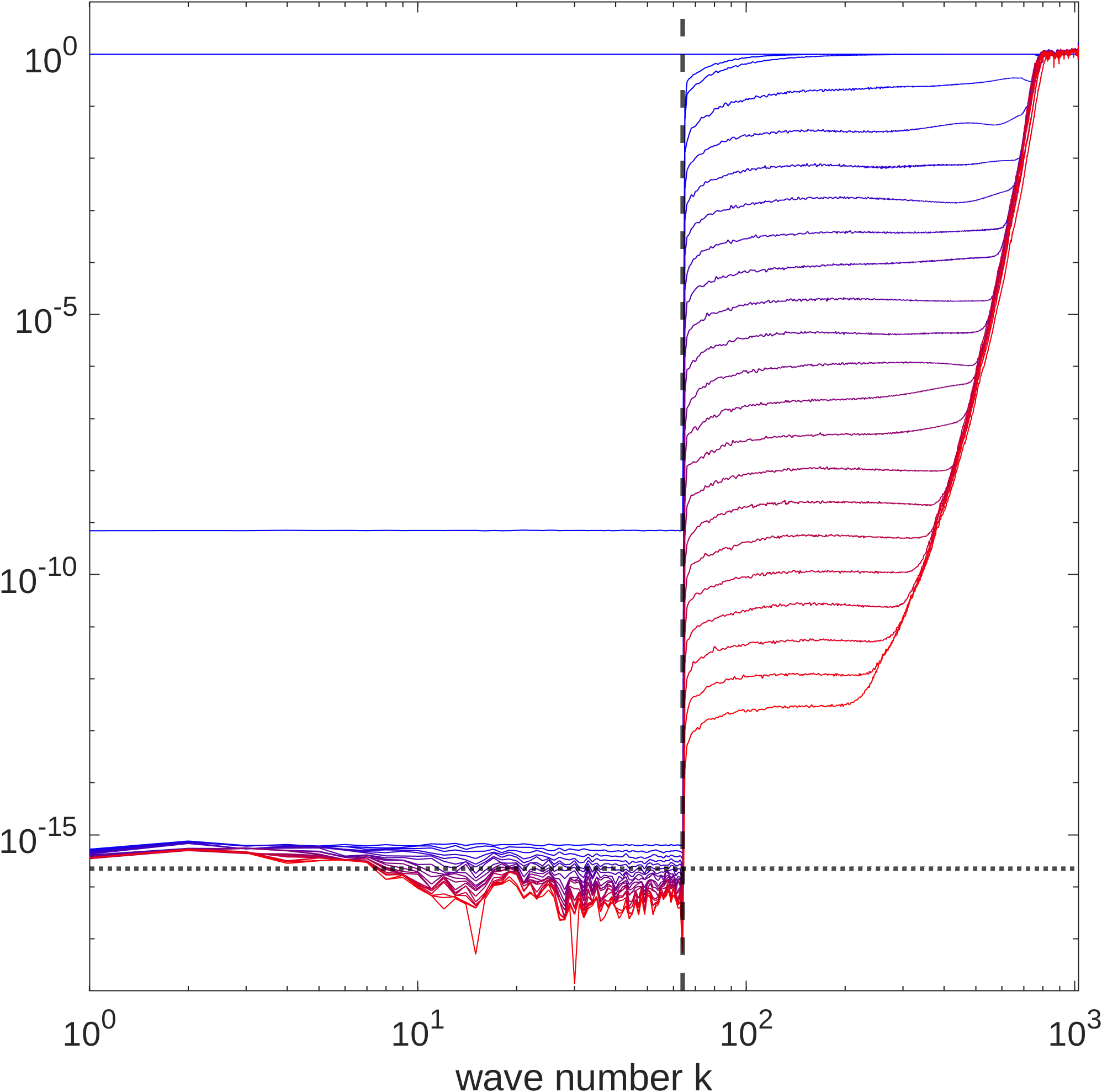}
\caption{Relative spectral error in solution: $E_0[\psi - \psi_{DA}](k)/E_0[\psi](k)$ vs. $k$.}
\end{subfigure}
\begin{subfigure}[b]{0.49\textwidth}
\includegraphics[width=1.0\linewidth,trim = 0mm 0mm 0mm 0mm, clip]{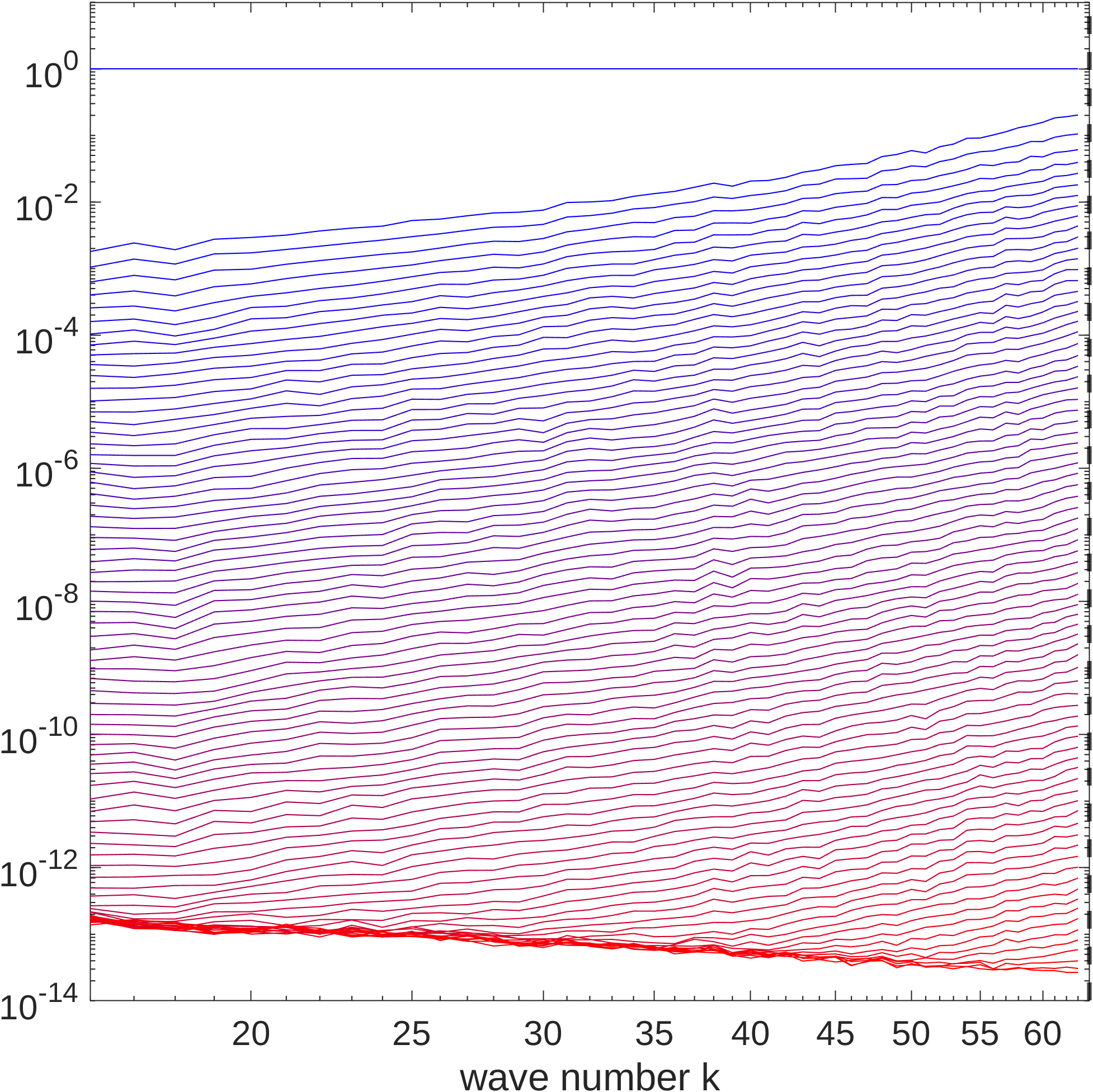}
\caption{Relative spectral error in forcing: $E_0[f^\psi - f^\psi_{DA}](k)/E_0[f^\psi](k)$ vs. $k$.}
\end{subfigure}
\caption{\label{fig:psi_err_spec.png} (log-log plot) Spectra\footnote{The reason for the difference in the $x$-axes is that the force is generated from only a projection onto the observational modes (namely, the modes $\mathbf{k}=(k_1,k_1)$ for which $|\mathbf{k}|_\infty:=\max(|k_1|,|k_2|)\leq64$).  In two dimensions, this means that for 
    $|\mathbf{k}|_2 = \sqrt{|k_1|^2+|k_2|^2}>64\sqrt{2}\approx 90.5$, the $\mathbf{k}^{\text{th}}$ modes will be zero.  Since $\log(0)$ is undefined, these do not show up on a log-log plot.} 
    of the error in the stream function (a) and the error of the forcing function (b) at times $t=0.0, 0.5, 1.0$, and also $t=2, 4, 6\ldots,40$ using direct replacement scheme. Colors move from blue ($t=0.0$) to red ($t=40.0)$. Vertical dashed line is the observational wave-number cut-off at $|\mathbf{k}|_\infty=64$.  Horizontal dotted line is machine precision ($\epsilon\approx2.22\times10^{-16}$). (c) Relative error in spectrum spectrum at times $t=0.0, 0.5, 1.0$, and also $t=2, 4, 6\ldots,80$ using direct replacement scheme.(d) Relative error in forcing spectrum at times $t=0.0, 0.5, 1.0$, and also $t=2, 4, 6\ldots,80$ using direct replacement scheme.}
% NOTE (Don't delete!): This is data from NSE_2048_G2500000_nu0p0001_alpha0_damp0_1660807556
\end{figure}

In addition to variations in the forcing update, we also updated the forcing continuously (or at every discrete time step in the simulation) or at uniformly-spaced time intervals.  In all cases, the force and the solution converged exponentially fast with very little distinguishing differences amongst the different simulations, as shown in FIG. \ref{fig:b}.  We anticipate this is because the relaxation time scale required between forcing updates is akin to $1/\mu$ which for the selected values of $\mu$ is on the same order as the actual time step of the numerical algorithm so that `continuous' updating of the forcing is on the requisite time scale anyway. We also observed the time-evolution of the convergence of the energy spectrum of the errors of the stream function and the forcing function (FIG. \ref{fig:psi_err_spec.png}).  These figures demonstrate that the state (streamfunction in this case) converges on the observable scales (below the cutoff of $|\mathbf{k}|_\infty = 64$) almost instantaneously, but the unobserved scales of the state, and the full forcing function (which is everything in the observed range) converge much slower.
% psi: 1863224 unknowns
% \begin{figure}[ht]
% \includegraphics[width=1.0\linewidth,trim = 0mm 0mm 0mm 0mm, clip]{init_data.png}
% \caption{\label{fig:a} Initial vorticity. Resolution $2048^2$.}
% \end{figure}
%
% \begin{figure}[ht]
% \includegraphics[width=1.0\linewidth,trim = 0mm 0mm 0mm 0mm, clip]{force.png}
% \caption{\label{fig:a} Forcing function. Resolution $2048^2$.}
% \end{figure}

% \begin{figure}[ht]
% \includegraphics[width=1.0\linewidth,trim = 0mm 0mm 0mm 0mm, clip]{2048_convergence_L2_g_force_error.jpg}
% \caption{\label{fig:b} $L^2$ error in the curl of the force vs. time.}
% \end{figure}

% \begin{figure}[ht]
% \includegraphics[width=1.0\linewidth,trim = 0mm 0mm 0mm 0mm, clip]{2048_convergence_L2_psi_error.jpg}
% \caption{\label{fig:c} $L^2$ error in the stream function vs. time.}
% \end{figure}

% \begin{figure}[ht]
% \includegraphics[width=1.0\linewidth,trim = 1mm 7mm 14mm 13mm, clip]{psi_err_spec.jpg}
% \caption{\label{fig:psi_err_spec.png} (log-log plot) Spectrum of the error in the stream function at times $t=0.0,0.5\ldots,40.0$. Colors move from blue ($t=0.0$) to red ($t=40.0)$. Vertical dashed line is the observational wave-number cut-off at $|k|=64$.  Horizontal solid line is machine precision ($\epsilon\approx2.2\times10^{-16}$).}
% % NOTE (Don't delete!): This is data from NSE_2048_G2500000_nu0p0001_alpha0_damp0_1660807556
% \end{figure}

% \begin{figure}[ht]
% \includegraphics[width=1.0\linewidth,trim = 1mm 7mm 14mm 13mm, clip]{force_err_spec.jpg}
% \caption{\label{fig:force_err_spec.png} (log-log plot) Spectrum of the error in the stream function forcing at times $t=0.0,0.5\ldots,40.0$. Colors move from blue ($t=0.0$) to red ($t=40.0)$. Vertical dashed line is the observational wave-number cut-off at $|k|=64$.  Horizontal solid line is machine precision ($\epsilon\approx2.2\times10^{-16}$).}
% % NOTE (Don't delete!): This is data from NSE_2048_G2500000_nu0p0001_alpha0_damp0_1660807556
% \end{figure}

% \begin{figure}[ht]
% \includegraphics[width=0.49\textwidth,trim = 0mm 0mm 0mm 0mm, clip]{force_error_spec_DR_trim.png}
% \caption{\label{fig:force_err_spec.png} (log-log plot) Spectrum of the error in the stream function forcing at times $t=0.0,0.5\ldots,40.0$ using direct replacement scheme. Colors move from blue ($t=0.0$) to red ($t=40.0)$. Vertical dashed line is the observational wave-number cut-off at $|k|=64$.  Horizontal solid line is machine precision ($\epsilon\approx2.2\times10^{-16}$).  The difference in the horizontal scale from the previous Figure is due to the lack of information in the higher modes for the true forcing function $\mathbf{f}$.}
% % NOTE (Don't delete!): This is data from NSE_2048_G2500000_nu0p0001_alpha0_damp0_1660807556
% \end{figure}

%\vspace{1in}

\section{Conclusions}\label{sec:conclusions}

In conclusion, we have developed a novel algorithm for recovering an unknown large scale forcing function in the 2D Navier-Stokes equations that can be implemented on the fly. The algorithm is both rigorously and numerically justified, simple to implement, and does not require an ensemble of simulations to generate the desired forcing function. In particular, the algorithm here will be applicable to a host of other situations where dissipative mechanisms and driving forces are simultaneously present; this is typically the case in many hydrodynamic settings, especially ones arising in the context of climate and weather modeling.  We emphasize that the rigorous result provided here is restricted to two dimensions only due to the current lack of a global-in-time bound on solutions for the gradient of the velocity field for the three dimensional Navier-Stokes equations.  Under suitable assumptions on the regularity of solutions to the 3D equations, we are confident that the same algorithm presented here will recover the forcing in that setting.  Hence, we do not anticipate that the recovery of the forcing critically relies on the inverse cascade in 2D turbulence which tends to aggregate large scale structures \cite{boffetta2012two}.  Further investigations (both numerical and analytical) are required to confirm these conjectures however.

\FloatBarrier

\begin{acknowledgments}
We wish to acknowledge the ADAPT group and J. Murri and B. Pachev in particular who invigorated discussions surrounding these results developed here. 
A.F. was partially supported by NSF grant DMS-2206493. A.L. was supported in part by NSF grants DMS-2206762 and CMMI-1953346, and USGS  grant G23AS00157 number GRANT13798170.  V.M. was partially supported by NSF grants DMS-2206491 and DMS-2213363. J.P.W. was partially supported by NSF grant DMS-2206762, as well as the Simons Foundation travel grant under 586788.
\end{acknowledgments}

\FloatBarrier

%\appendix

\section{Appendix}\label{sec:appendix}

We provide some additional supporting details for the convergence analysis performed after \eqref{eq:error:rep}.
In particular we utilize several inequalities in the rigorous analysis which are described below.  First we note that we will make use of the $L^p$ norm on functions defined for functions $\phi:\Omega \rightarrow \mathbb{R}$ on the domain $\Omega = [0,L]^2$ as
\begin{equation*}
    \|\phi\|_p = \left(\int_\Omega |\phi(\mathbf{x})|^p d\mathbf{x}\right)^{1/p},
\end{equation*}
with $p=\infty$ corresponding to the supremum norm over the entire domain.

This allows us to describe the following inequalities defined for all functions $\phi,\psi,\chi:\Omega \rightarrow\mathbb{R}$ (extension to vector-valued functions is immediate):
\begin{itemize}
    \item Cauchy-Schwarz inequality:
    \begin{equation}
        \int_\Omega |\phi(\mathbf{x})\psi(\mathbf{x})| d\mathbf{x} \leq \|\phi\|_2\|\psi\|_2.
    \end{equation}
    \item Agmon interpolation inequality:
    \begin{equation}
        \|\phi\|_\infty \leq c \|\nabla \phi\|_2^{1/2}\|\nabla^2\phi\|_2^{1/2},
    \end{equation}
    where $c$ is a universal constant.
    \item Ladyzhenskaya interpolation inequality:
    \begin{equation}
        \|\phi\|_4 \leq c \|\nabla \phi\|_2^{1/2}\|\phi\|_2^{1/2},
    \end{equation}
    where once again $c$ is a universal constant.
    \item Generalized H\"older inequality:
    \begin{equation}
        \|\phi \psi \chi\|_r \leq \|\phi\|_p \|\psi\|_q \|\chi\|_s,
    \end{equation}
    where $\frac{1}{s}+\frac{1}{q}+\frac{1}{p} = \frac{1}{r}$.
    \item Gr\"onwall inequality:
    
    Suppose that $f(t),g(t):[0,T]\rightarrow\mathbb{R}$ are continuously differentiable, and satisfy
    \begin{equation}
        \frac{df}{dt} \leq \alpha f(t) + g(t),
    \end{equation}
    for some constant $\alpha$.  Then
    \begin{equation}
        f(t) \leq f(0) e^{\alpha t} + \int_0^t e^{\alpha (t-s)}g(s)ds.
    \end{equation}
\end{itemize}

The rigorous analysis also makes use of \textit{a priori} bounds on the velocity field corresponding to \eqref{eq:NS_true}. To state these bounds, suppose $\mathbf{f},\nabla\mathbf{f}\in L^2(\Omega)$ such that $\mathbf{f}$ is divergence-free. Let $\kappa_0$ and $G$ be defined by \eqref{Grashof} and denote a \textit{shape factor} of the force by
    \[
        \sigma_1:=\frac{\kappa_0\|\nabla\mathbf{f}\|_{2}}{\|\mathbf{f}\|_{2}}.
    \]

Strong solutions of \eqref{eq:NS_true} satisfy
    \begin{align}\label{est:apriori}
        \begin{split}
    \|\mathbf{u}(t)\|_{2}^2&\leq \|\mathbf{u}_0\|_{2}^2e^{-\kappa_0^2\nu t}+\kappa_0^2\nu^2G^2 < R_0\\
    \|\nabla\mathbf{u}(t)\|_{2}^2&\leq \|\nabla\mathbf{u}_0\|_{2}^2e^{-\kappa_0^2\nu t}+\nu^2G^2 < R_1\\
    \|\Delta\mathbf{u}(t)\|_{2}^2&\leq \|\Delta\mathbf{u}_0\|_{2}^2e^{-\kappa_0^2\nu t}+c\kappa_0\nu^2 (\sigma_1+G)^2G^2 < R_2,
        \end{split}
    \end{align}
for all $t\geq0$. The first two bounds are classical and may be found in \cite{ConstantinFoias88, Temam1997, FoiasManleyRosaTemamBook2001}, whereas the third can be found in, for instance, \cite{martinez2022force}; we have taken the $R_j$ to be strict upper bounds on these estimates. 
Note that when $t$ is sufficiently large, one may assume that $R_0, R_1, R_2$ are independent of the initial velocity; this is a safe assumption in the analysis performed above since we must always wait long enough before the first update to ensure that $\mathbf{u}(t), \Delta\mathbf{u}(t)$ henceforth remain bounded by $R_0$ and $R_2$, respectively.

Lastly, the crucial ingredient used to close all of the estimates and arrive at \eqref{eq:pre:convergence} is provided by the state error estimates \eqref{eq:error:sync:true}. The bound stated for $\nabla\mathbf{w}^k$ was originally proven in \cite{martinez2022force}. We state its precise form here: Let $\mathbf{u}_0,\mathbf{v}_0$ be divergence-free velocity fields, which are mean-free over $\Omega$, and whose derivatives up to second-order are square-integrable over $\Omega$. Then there exist universal constants $\overline{c}_0, \underline{c}_0\geq1$ such that if $\mu, N$ satisfy
    \begin{align}\underline{c}_0\nu\left(\sigma_1+G\right)G^2\leq {\mu h^2}\leq \overline{c}_0,\notag
    \end{align}
then for each $k\geq1$, there exists $t_k>t_{k-1}>0$ (where $t_0:=0$), such that
    \begin{align}
        \|\nabla\mathbf{w}^k(t)\|_{2}
        &\leq \left(\frac{2C_1\kappa_0^2\nu}{\mu}\right)^{1/2}\frac{\|\Delta\mathbf{g}^{k}(t_k)\|_{2}}{\kappa_0\nu},\notag
    \end{align}
for all $t\in [t_k,\infty)$, for some universal constant $C_1\geq1$ independent of $k$.

On the other hand, the bound for $\mathbf{w}^k$ was not required in \cite{martinez2022force}. For the sake of completeness, we provide the details of this estimate here. Observe that the system governing the evolution of $\mathbf{w}^k=\mathbf{v}^k-\mathbf{u}$ is given by
    \begin{align}
        \partial_t\mathbf{w}^k-&\nu\nabla^2\mathbf{w}^k+\mathbf{w}^k\cdotp\nabla\mathbf{w}^k+\mathbf{w}^k\cdotp\nabla\mathbf{u}+\mathbf{u}\cdotp\nabla\mathbf{w}^k=-\nabla r^k+\Delta\mathbf{g}^k-\mu I_h\mathbf{w}^k,\quad\nabla\cdotp\mathbf{w}^k=0\notag
    \end{align}
where $r^k=q^k-p$, where $q^k, p$ denote the pressures corresponding to $\mathbf{v}^k,\mathbf{u}$, respectively. Then the corresponding energy balance is given by
    \begin{align}
        \frac{1}2&\frac{d}{dt}\|\mathbf{w}^k\|_{2}^2+\nu\|\nabla\mathbf{w}^k\|_{2}^2+\mu\|\mathbf{w}^k\|_{2}^2=-\langle \mathbf{w}^k\cdotp\nabla\mathbf{u},\mathbf{w}^k\rangle+\langle\Delta\mathbf{g}^k,\mathbf{w}^k\rangle+\mu\|J_h\mathbf{w}^k\|_{2}^2\notag.
    \end{align}
The first of these terms on the right-hand side can be estimated with H\"older's inequality, Ladyzhenskaya's inequality, the Cauchy-Schwarz inequality, and \eqref{est:apriori} to obtain
    \begin{align}
        |\langle \mathbf{w}^k\cdotp\nabla\mathbf{u},\mathbf{w}^k\rangle|\leq\|\nabla\mathbf{u}\|_2\|\mathbf{w}^k\|_{4}^2\leq cR_1\|\nabla\mathbf{w}^k\|_2\|\mathbf{w}^k\|_2\leq c\frac{R_1^2}{\mu}\|\nabla\mathbf{w}^k\|_2^2+\frac{\mu}{4}\|\mathbf{w}^k\|_2^2\notag.
    \end{align}
The second term can be estimated with the Cauchy-Schwarz inequality to obtain
    \begin{align}
        |\langle\Delta\mathbf{g}^k,\mathbf{w}^k\rangle|\leq\|\Delta\mathbf{g}^k\|_2\|\mathbf{w}^k\|_2\leq\frac{1}{\mu}\|\Delta\mathbf{g}^k\|_2^2+\frac{\mu}4\|\mathbf{w}^k\|_2^2.
    \end{align}
Lastly, the third term can be estimated using the Cauchy-Schwarz inequality and \eqref{eq:boundedness} to obtain
    \begin{align}
        \mu\|J_h\mathbf{w}^k\|_2^2\leq C_1^2\mu h^2\|\nabla\mathbf{w}^k\|_2^2\notag.
    \end{align}
Now let us assume that $\mu,h$ satisfy
    \[
    2C_1^2\mu h^2\leq \nu,\quad c\mu\nu\geq R_1^2.
    \]
We may then combine the above estimates to arrive at
    \begin{align}
        \frac{d}{dt}\|\mathbf{w}^k\|_2^2+\mu\|\mathbf{w}^k\|_2^2\leq \frac{1}{\mu}\|\Delta\mathbf{g}^k\|_2^2,\notag
    \end{align}
from which \eqref{eq:error:sync:true} follows upon taking $t$ sufficiently large to evaluate $\mathbf{g}^k$, and finally an application of Gr\"onwall's inequality.

%\nocite{*}

\bibliographystyle{abbrv}
\providecommand{\noopsort}[1]{}\providecommand{\singleletter}[1]{#1}%

% Produces the bibliography via BibTeX.

\end{document}